\documentclass[10pt, conference, letterpaper]{IEEEtran}\IEEEoverridecommandlockouts
\usepackage{cite}
\usepackage{amsmath,amssymb,amsfonts}
\usepackage{algorithmic}
\usepackage{graphicx}
\usepackage{textcomp}
\usepackage{xcolor}
\usepackage{url}

\usepackage{url}
\usepackage{makecell}
\usepackage{subcaption}
\usepackage{amsmath}
\usepackage{pifont}% http://ctan.org/pkg/pifont
\usepackage{times}  
\usepackage{hyperref}

\def\BibTeX{{\rm B\kern-.05em{\sc i\kern-.025em b}\kern-.08em
    T\kern-.1667em\lower.7ex\hbox{E}\kern-.125emX}}
\begin{document}

%\begin{frontmatter}

\title{Bounding Queue Delay in Cellular Networks to Support Ultra-Low Latency Applications}

%\author{Soheil Abbasloo}
%\author{H. Jonathan Chao}
%\address{ab.soheil@nyu.edu, chao@nyu.edu}
%\address{New York University, New York, USA}

\author{\IEEEauthorblockN{Soheil Abbasloo}
\IEEEauthorblockA{%\textit{Tandon School of Engineering} \\
\textit{New York University}\\
ab.soheil@nyu.edu}
\and
\IEEEauthorblockN{Jonathan H. Chao}
\IEEEauthorblockA{%\textit{Tandon School of Engineering} \\
\textit{New York University}\\
chao@nyu.edu}
}

\maketitle

\begin{abstract}
Most of the current active queue management (AQM) designs have major issues including severe hardship of being tuned for highly fluctuated cellular access link bandwidths. Consequently, most of the cellular network providers either give up using AQMs or use conservative offline configurations for them. However, these choices will significantly impact the performance of the emerging interactive and highly delay sensitive applications such as virtual reality and vehicle-to-vehicle communications. 

Therefore, in this paper, we investigate the problems of existing AQM schemes and show that they are not suitable options to support ultra-low latency applications in a highly dynamic network such as current and future cellular networks. Moreover, we believe that achieving good performance does not necessarily come from complex drop rate calculation algorithms or complicated AQM techniques. Consequently, we propose BoDe an extremely simple and deployment friendly AQM scheme to bound the queuing delay of served packets and support ultra-low latency applications. 

We have evaluated BoDe in extensive trace-based evaluations using cellular traces from 3 different service providers in the US and compared its performance with state-of-the-art AQM designs including CoDel and PIE under a variety of streaming applications, video conferencing applications, and various recently proposed TCP protocols. Results show that despite BoDe's simple design, it outperforms other schemes and achieves significantly lower queuing delay in all tested scenarios.
\end{abstract}


%\end{frontmatter}
%
%\linenumbers

\section{Introduction}
\label{sec_intro}
%From a historical point of view, the delay tolerant (and in most cases delay insensitiveness) nature of the current/traditional applications and the fact that there is always a trade-off between minimizing overall delay and maximizing overall throughput have pushed a majority of the network protocol/system designs toward being throughput-oriented structures. A simple example is the dominance of TCP Cubic\cite{cubic}, a loss-based throughput-oriented TCP design, as the default TCP scheme in most of today’s smartphone and PC platforms.
Emerging interactive and highly delay sensitive applications such as virtual reality and vehicle-to-vehicle communications impose new stringent delay requirements on the future cellular networks. However, most of the current network system/protocol designs are mainly rested on the traditional throughput-oriented structures. This sheds light on the need for rethinking about network system/protocol designs to satisfy new end-to-end (e2e) ultra-low latency requirements. Considering that, recently, vast amount of studies focused on the delay-centric designs either for cellular or wired networks (e.g. \cite{c2tcp,c2tcp2}, \cite{sprout}, \cite{vivace}). Due to the fact that usually fully e2e solutions are more deployment friendly than solutions requiring a change in the network (in-network solutions), nearly all of these new studies propose an e2e solution for reducing the overall delay of the packets. 

However, one of our key statements is that no matter how good an e2e solution is, it cannot alone satisfy ultra-low delay demands in a highly dynamic network such as the cellular network. The main insight is that the help of the network and especially having a proper AQM design for the cellular network queues is needed to meet stringent delay requirements of highly interactive and real-time applications (detailed in section \ref{sec_e2e}). 

%QoS in 3GPP spec defined a high-level specification for various QCI traffic such as..... These specs are mainly proposed to --> have multivendor alignment/... -> with the main intention of directing the scheduling functionality among various bearer for different providers. 

Interestingly, to the best of our knowledge, most of the cellular network providers do not use AQM designs. Instead, they leverage deep buffers at base stations (BTS) to avoid/minimize drop of packets in their networks and achieve high reliability. 

The two main reasons behind that are 1) the serious tuning issues of current AQM schemes and 2) the throughput-oriented nature of current applications. 
\begin{itemize}
\item \textbf{The Tuning Issues}: A rule of thumb for classic AQMs is to set the network's buffer sizes to bandwidth-delay product (BDP) of the network. However, cellular networks experience very high variations of the wireless link's bandwidth. So, a good buffer size setting for a specific wireless link bandwidth will lead to poor performance when the capacity of the wireless link changes. Hence, it is very difficult (if not impossible) to have a fixed pre-configured parameter setting for most AQMs so that they can perform well in all different cellular network conditions. 

\item \textbf{The Throughput-Oriented Applications}: Today's applications are mainly throughput-oriented applications. Therefore, the end-users' satisfaction will not be impacted that much if network operators either use no AQM schemes or use conservatively tuned parameters for the AQM schemes. 
\end{itemize}

Consequently, most of the cellular network providers either give up using AQM schemes or use extremely conservative offline configurations for them (for instance, they consider \textit{maximum} BDP of the network to set buffer sizes). However, these choices will impact the performance of emerging ultra-low latency applications.
	
New AQM designs including state-of-the-art ones such as PIE\cite{pie} and CoDel\cite{codel} try to tackle the tuning issues of classic AQMs. However, even these schemes have design issues which make them suboptimal solutions. The main issue of them is their limited \textit{scope of applicability}. For instance, one of the fundamental assumptions of both PIE and CoDel (and nearly all of the AQM schemes in the literature) is that the sender uses a \textit{loss-based TCP} to send its traffic to the network. Loss-based TCP is a specific category of congestion control designs in which the loss of packets in the network is considered as the only signal of congestion (e.g. ~\cite{cubic,tahoa,newreno,bic}). However, exploiting UDP-based designs (which use their own algorithms to control the sending rate) for current interactive applications such as Skype (and most likely for the future ones), QUIC\cite{quic} used by Google on Chrome browser, and a lot of recent new trends in transport control design, including using machine learning techniques (e.g. \cite{vivace} and \cite{remy}), congestion-based designs (e.g. \cite{bbr}), and delay-based designs (e.g. \cite{vegas} and \cite{c2tcp}) show why the assumption of having a loss-based TCP as the source of the traffic is not necessarily correct.

Inevitable need for having an AQM design and issues with the current AQM designs motivated us to propose a \textbf{Bo}unded \textbf{De}lay AQM called \textit{BoDe} to boost the performance of highly delay sensitive interactive applications in cellular networks which can work transparently with any algorithm or transport control mechanism running at the end-hosts. BoDe's logic is extremely simple and due to the recent advances in the programmable network devices~\cite{p4,%sw1,sw2,
sw3,sw4}, it is a deployment friendly approach. 

Using 4G cellular traces gathered in NYC and Boston by prior work~\cite{sprout,c2tcp}, we evaluated BoDe and compared it with various AQM schemes including state-of-the-art schemes such as CoDel\cite{codel} and PIE\cite{pie} and baseline schemes such as TailDrop and HeadDrop queues. To test BoDe's performance, we used variety of transport control designs including Sprout\cite{sprout}, BBR\cite{bbr}, Westwood\cite{west}, Cubic\cite{cubic}, PCC-Vivace\cite{vivace}, and C2TCP\cite{c2tcp}. In addition, we examined BoDe using Skype, Google Hangout, YouTube, AmazonPrime, and various adaptive bit rate (ABR) streaming protocols including BB\cite{bb}, FESTIVE\cite{festive}, BOLA\cite{bola}, and MPC\cite{mpc}. Results show that compared to the baseline schemes, BoDe can achieve more than $170\times$ lower 99th percentile queuing delay and compared to CoDel and PIE, it can reduce the 99th percentile queuing delay from $2\times$ to $20\times$ considering all tested scenarios. This great delay performance comes with a compromise in throughput. In the worst case, BoDe compromises throughput 40\% compared to the best throughout (achieved using TailDrop queue).

%BoDe's great performance indicates that in contrast with the tendency of cellular network service providers to achieve ultra-reliability and drop nearly no packets, dropping packets in the cellular network is an important part of the solution to support applications with ultra-low latency requirements.
\section{Related Work}
%\textbf{AQM schemes:}
Active queue management schemes deal with problems such as bufferbloat in the network itself. 
Schemes such as RED~\cite{red}, SRED~\cite{sred}, REM~\cite{rem}, BLUE~\cite{blue}, AVQ~\cite{avq}, and \cite{pi} use the idea of dropping/marking packets so that end-points can react to packet losses and control their sending rates. These AQM schemes detect congestion mainly based on the queue lengths (e.g., RED), the arrival rate of the packets (e.g.,~\cite{gibbens}), or a combination of both (e.g.,~\cite{pi}). However, tuning parameters of these schemes for a highly dynamic network such as cellular network usually is troublesome. Instead of controlling the queue length, recent schemes such as CoDel~\cite{codel} and PIE~\cite{pie} try to control the \textit{statistics} of the queuing delay. For instance, CoDel tries to keep the \textit{minimum} delay of the queue around a Target delay, while PIE uses the depletion rate of the queue to control the \textit{average} queuing delay and keep it around a reference point. However, the key assumption of all these schemes including the recent ones is to have a loss-based TCP as the source of traffic (detailed in Section~\ref{sec_aqm-tcp}). This throughput-oriented design structure cannot achieve good performance when used for ultra-low latency traffic (detailed in Section~\ref{sec_eval}). 

Explicit congestion notification (ECN) is another technique used to do the buffer management indirectly. In other words, ECN-based approaches, use modified switches to tag certain packets and explicitly notify the senders about congestion in the network. Later, senders need to adjust their sending rates to reduce queue occupancy and congestion in the network. ABC~\cite{abc} is a recent example of ECN-based approach proposed for cellular networks. However, there are two key issues with all ECN-based approaches. First, to employ these approaches in practice, application (or network-stack) at the client/server need to be modified to process the ECN bits set by the switches and change the sending rates accordingly. This means that for TCP-based applications, a Kernel patch at all sources/destinations is required. The problem becomes worst for UDP-based applications (such as Skype, QUIC~\cite{quic}etc.) that manage the congestion in the application layer. This means that all these UDP-based applications need to be modified to use the ECN-bits which is not a deployment friendly solution. The second issue is the fact that the network service providers are usually not interested in exposing their network's states (queue occupancy, queue delay, etc.) required by ECN-based schemes to the end-users. 

Recently, NATCP~\cite{natcp} proposes a new-fashioned design approach that uses direct feedback sent by the network to the end-hosts so that the end-hosts can control their sending rates and minimize the delay of their packets while maximizing the throughput. It is shown that NATCP can perform close to the optimal delay and throughput performance. However, its novel design approach requires adding new entity called NetAssist to the network.

These issues with the new approaches and the fact that conventional throughput-oriented design structures cannot achieve good performance for ultra-low latency traffic motivates us to rethink the AQM in cellular networks and propose a new simple yet effective delay-oriented AQM design to boost the performance of emerging ultra-low latency traffic without exposing any network information to and without requiring any changes at the end-hosts. 

\section{Setting the Context}
\label{sec_back}
To set the context, here, we briefly describe the cellular networks' unique characteristics and issues and properties of delay sensitive applications.

\textbf{Cellular Networks}: Cellular networks differ from the wired networks in several ways resulting in overall poor performance when protocols/algorithms designed originally for wired networks are used in cellular networks. Three of these main differences are: 
\begin{enumerate}
\item Per-UE large queues at BTS (base station): Having per-UE large queues provides a good structure to properly isolate the traffic of different UEs. However, this isolation brings a new form of delay issue called self-inflicted queuing delay\cite{sprout,c2tcp}. 
\item High link capacity fluctuations: Compared to the wired networks, cellular networks can experience multiple order of magnitude faster channel capacity variations.
\item Wireless scheduling: Wireless scheduler at BTS is responsible to manage fairness among different UEs by scheduling them on certain timeslots to access the wireless channels. So, even when the channel capacity is good for a certain UE, it might not be able to use it. Thus, the available link capacity of UEs is varying even more than their actual physical channel capacity. This phenomenon causes the packets of UEs to experience a so-called scheduling delay in both directions. 
\end{enumerate}
\begin{figure*}[!tp]
\centering
\includegraphics[width=0.9\linewidth,height=1.5in]{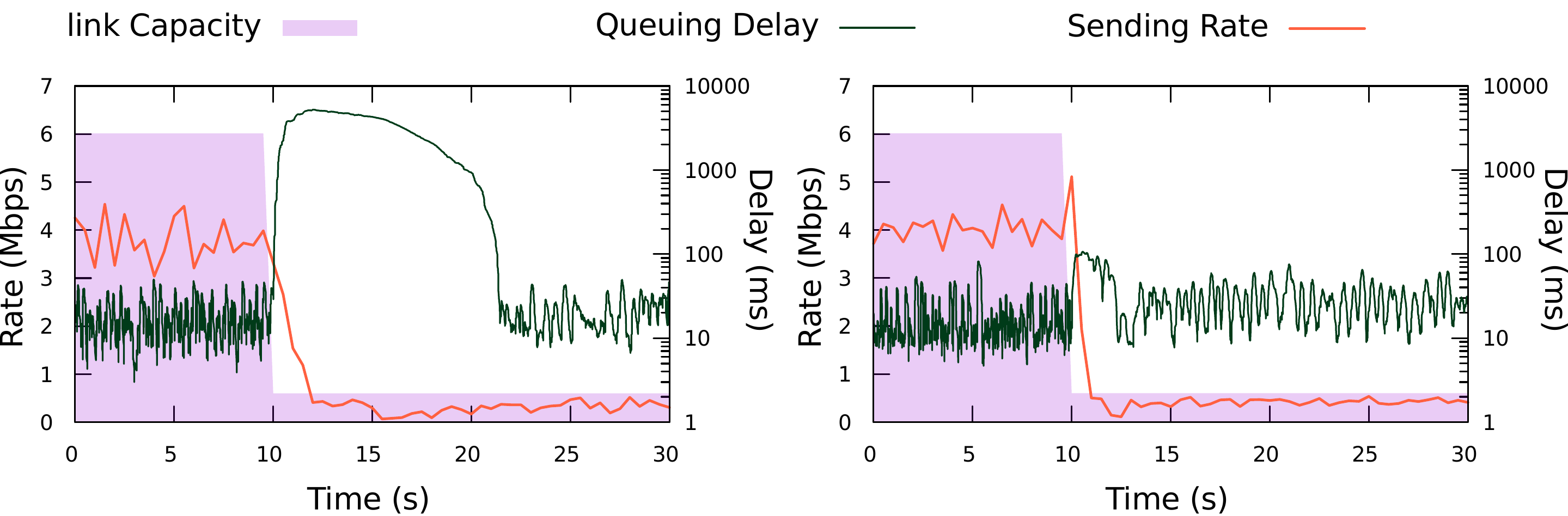}
\caption{Queuing delay and sending rate of Skype using FIFO queue (left) and BoDe queue (right) at BTS}
\label{fig_skype-step}
\end{figure*}
\textbf{Emerging Delay Sensitive Applications}: Emerging ultra-low delay-sensitive applications such as interactive VR, automated vehicles, etc. define new network requirements. The key difference of these interactive applications compared to traditional ones such as web applications is that newer packets are considered more important than older packets. For instance, in an interactive video call, video packets sent 3 seconds before can be ignored if this helps to receive the newer packets. However, in a file downloading application, there is no advantage of sacrificing older packets to speed up newer packets. All packets are equally important in such applications.
%\begin{figure*}[!t]
%    \centering
%    \begin{minipage}[b]{0.45\linewidth}
%        \centering
%            \includegraphics[width=0.8\linewidth,height=1.3in]{fifo-del-thr}
%    \end{minipage}
%    \begin{minipage}[b]{0.45\linewidth}
%            \centering
%        \includegraphics[width=0.8\linewidth,height=1.3in]{bode-del-thr}
%    \end{minipage}
%\caption{Queuing delay and sending rate of Skype using FIFO queue (left) and BoDe queue (right) at BTS}
%\label{fig_skype-step}
%\end{figure*}

\textbf{Emerging Architectures and Server Proximity}: To satisfy new demands of these new applications, new architectures such as MEC (mobile/multi-access edge computing)\cite{mec} and MCDN (mobile content delivery network) have been introduced to push delay-sensitive contents close to the UEs. Although these trends greatly reduce the intrinsic e2e latency, due to the issues described earlier in cellular networks, wireless access links (known as last-mile) remain the main bottleneck link in cellular networks. Therefore, considering server proximity, we assume last-mile is the bottleneck link and the source of the main delay throughout this paper. 

\section{The Design, Part I: Motivations and Design Decisions}
In this section, we explain the main design decisions behind BoDe. More specifically, we illustrate why an AQM scheme in the network is required for supporting ultra-low latency applications. Then, we discuss the general issues with existing AQM schemes in detail. 
\subsection{Why Not Just Another e2e Design?}
\textbf{Q1: Can fully e2e schemes meet ultra-low latency and high throughput requirements in cellular networks?}
\label{sec_e2e}
To answer this important question which impacts the overall design direction, we focus on the link capacity fluctuations of the last mile caused by either wireless scheduler or change in the quality of the channel for UE. These fluctuations can be boiled down into 2 cases: Case 1- When available capacity suddenly increases, Case 2- When available capacity suddenly decreases. In the best scenario, e2e schemes can handle Case 1 by increasing their sending rates. However, considering delay and throughput, no fully e2e scheme can perform well under Case 2. To describe the reason, we do a simple experiment. We use a step function to emulate a sudden change in cellular access link capacity (similar to case 2), use FIFO queue at BTS, and send Skype traffic through this link as shown in Fig.~\ref{fig_skype-step} (the left graph)\footnote{Check section \ref{sec_inter} for more details of the evaluation setup}. Before the sudden drop of capacity (at time 10), the sender already sent some number of packets to the network assuming that the link capacity is still 6Mbps. Therefore, at time 10 when the link capacity becomes 0.6Mbps, there will be a sudden self-inflicted queue buildup (up to 4 seconds). At this time, the sender cannot do anything to reduce the number of in-flight packets that are already sent to the network even in the best case where the sender can be notified immediately by the network. However, if the network itself could have been allowed to manage those overshoots using AQM schemes such as BoDe, these self-inflicted queue buildups, caused by the packets belonging to the past, would have been resolved (see right graph in Fig.~\ref{fig_skype-step}). 

Although self-inflicted queue buildups won't impact web-based applications (or in general, any traditional throughput hungry application), they dramatically degrade the performance of highly interactive and delay sensitive applications. 

%The key difference is that for an interactive application, current packets corresponding to the current time always have higher priority than the packets corresponding to the past. For instance, if the application is a video conference call, packets carrying the present's frames/video will be queued behind all packets belonging to the past (times smaller than $t_0$) and this self-inflicted queuing delay hugely impacts the interactivity nature of the application.

Therefore, our answer to the question Q1 is "No" and this leads to our main statement that \textit{using proper AQM designs and dropping packets in the cellular network are an important part of the solution to support future ultra-low latency applications}. 

\subsection{AQMs \& Inaccurate Assumptions}
\label{sec_aqm-tcp}
Nearly all of the current AQM designs are based on assumptions that are inaccurate in our context, i.e., ultra-low latency applications in cellular networks. The vast majority of AQM designs are based on the models which are described by Markov Chains (e.g. M/D/k, M/M/k, M/G/k, etc.) in which the key assumption is to have uncorrelated arrival processes. This key assumption might be a good approximation for a general queue on the Internet where there are different uncorrelated independent flows going to the same queue. However, this is not the case in the cellular network where a dedicated per-UE queue absorbs the traffic for only that UE and the arrival traffic is being controlled in a closed-loop fashion by using either an algorithm in the application layer (e.g., \cite{bb} and \cite{bola}) or in transport layer such as TCP. This closed-loop nature makes a complete correlated arrival process which is in contrast with the mentioned assumption.

Even when AQM designs consider this closed-loop nature of the traffic (e.g. as in \cite{codel} and \cite{pie}) they still rest on another assumption that the source of the traffic is a loss-based TCP (such as \cite{newreno},\cite{reno},\cite{cubic}, etc.) with a specific algorithm to deal with loss of packets. However, recent new trends in TCP design, including using machine learning techniques(e.g. \cite{vivace} and \cite{remy}), congestion-based designs (e.g. \cite{bbr}), delay-based designs (e.g. \cite{vegas} and \cite{c2tcp}) and use of UDP-based designs in interactive applications (such as Skype) which use their own proprietary algorithm to control the sending rate, and even use of QUIC\cite{quic} for streaming video by Google, shows having a loss-based TCP as the source of traffic is not necessarily a valid assumption anymore.

These inaccurate assumptions lead to performance degradation of AQM schemes for supporting ultra-low latency applications in cellular networks (detailed in Section~\ref{sec_eval}).

\section{The Design, Part II: BoDe}
BoDe's design rests on two simple principles: 1- In a highly dynamic network, making a decision for an incoming packet at the time of enqueuing it is not the best choice. 2- Instead of operating close to the globally optimal point, it is practically sufficient to operate around the applications' desired point.
\subsection{Live the Future; Then Make the Decision!}
\label{sec_future}
Assume that for a given cellular network, current queue length for a user is 100 packets (equal to current BDP (bandwidth delay product) of the network) and the objective is to reduce the overall delay experienced by packets.
\\
\textbf{Q2: What should an AQM scheme do to an incoming packet? Should it drop the packet?}
Actually, the answer depends on the \textit{future} state of the network! For instance, if the available link capacity in the next time slots doubles, keeping the 101st packet in the queue is a better choice. However, if the capacity halves in the next time slots, the better choice is to drop the packet. This simple question shows why due to the highly dynamic nature of the cellular network, making a decision for an incoming packet at the time of enqueuing it is not the best choice.

To address this issue, BoDe absorbs the impact of the \textit{future} changes of the available link on the current incoming packet, by postponing the drop decision from the time of enqueuing the packet to the time of dequeuing it. We will describe the details in Section~\ref{sec_alg}.

\subsection{Sufficient; Instead of Optimal}
\label{sec_optimal}
From the mathematical point of view, the operation point where the delay is minimized and throughput is maximized is the best~\cite{optimal}. However, J. Jaff~\cite{jaf} proved that this operation point cannot be reached by a distributed algorithm let alone a local mechanism in only one queue, i.e., AQM. Therefore, the goal of achieving a globally optimal point is not practical.  Fortunately, in practice, for any application including highly delay sensitive ones, there is a safe region of operation where end-users will be satisfied. Therefore, instead of the globally optimized solution, we seek a practically acceptable solution. 

For a delay sensitive application, any packet is required to reach the destination before a specific delay. For interactive applications, this deadline is determined by the nature of the application itself. For instance, delays in the order of 100ms for a voice/video signal cannot be detected by humans (it is a conventional standard interactivity delay). Another example is 20ms delay restriction for VR devices\cite{ar_vr} which means that the time between when the user changes its head direction and when the user sees the new frame (corresponding to the new field of view) should be less than 20ms to have a smooth transition. Considering that, there is no benefit in continuing to send a packet that has already passed its application's tolerable delay in the network.

Therefore, instead of serving all packets, BoDe considers the queuing delay of packets and bounds the sojourn time (waiting time) of them to a target called \textit{BoundedDelay}.  
\subsection{Algorithm}
\label{sec_alg}
BoDe's algorithm is simple and straightforward:
\begin{enumerate}
\item At ingress, tag arrival time of the packet (arrival time) 
\item At egress, serve packet if its sojourn time (= current time-arrival time) is smaller than \textit{BoundedDelay} of the queue. Otherwise, drop it.
\item If the queue length is smaller than 3 packets, do not drop the packet even if its sojourn time is larger than \textit{BoundedDelay}.
\end{enumerate}

Steps 1 and 2 are direct results of sections \ref{sec_future} and \ref{sec_optimal}. These steps will guarantee that no packet that has been served by BoDe in the queue experiences queuing delay more than \textit{BoundedDelay}. There is only one exception to this which is step 3. On the one hand, this exception comes from the fact that there is no benefit in dropping last packet(s) in the queue, because serving this packet(s) won't increase the sojourn time of any packets in the queue. On the other hand, this helps schemes with loss detection logic similar to TCP (such as detecting duplicate acks to trigger a fast retransmission)\footnote{Choosing 3 as the number of packets is to make sure that sender receives enough duplicate acks to trig mechanisms such as fast retransmission}. 

\textbf{How Does It Help Ultra-Low Latency Applications?}
In a nutshell, BoDe helps ultra-low latency applications in two key different ways. First, by bounding the queuing delay, BoDe assures applications that if they receive a packet, that packet has experienced a bounded delay in the network. Second, by dropping the packets experiencing large delays, BoDe indirectly informs applications about the network issues (congestion, bad link quality, scheduling delay etc.). This helps applications controlling their sending rates in a closed-loop fashion adjust their sending rates in a timely manner and prevent further drops of their packets.
\begin{figure*}[!t]
\centering
\includegraphics[width=0.95\linewidth,height=1.3in]{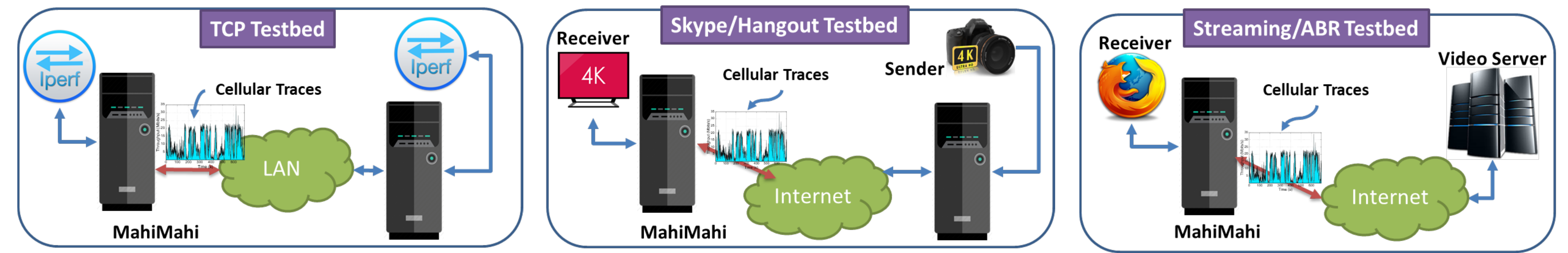}
\caption{Evaluation testbeds}
\label{fig_topo}
\end{figure*}
%\begin{figure}[!t]
%\centering
%\includegraphics[width=0.9\linewidth,height=1.5in]{topo}
%\caption{Evaluation testbeds}
%\label{fig_topo}
%\end{figure}

\textbf{What is the difference between BoDe and a simple head-drop FIFO queue?}
BoDe differs fundamentally from a head-drop queue. In head-drop's algorithm, drop of a HOL (head-of-line) packet happens in order to make room for accepting the new incoming packet. However, no incoming packet causes any drop of HOL packet in BoDe. Instead, BoDe's main philosophy is that making any drop decision (either dropping the incoming packet or dropping HOL packet) at the time of arriving an incoming packet is wrong. Considering the highly dynamic nature of cellular networks, BoDe gives a chance to the new incoming packet by letting it enter the queue and experience all future link capacity fluctuations (either capacity increase or capacity decrease). Then, at the time of serving that packet, it checks whether the given chance to the packet has led to an acceptable queuing delay for that packet.

\section{Evaluation} 
\label{sec_eval}
Here, we evaluate the performance of BoDe using real cellular traces and in a reproducible environment. We use Mahiamahi\cite{mahi} as our cellular emulator. We have implemented BoDe as a new queue structure in Mahimahi and use that implementation throughout this paper\footnote{The source code is available at: \url{https://github.com/Soheil-ab/bode.git}}. Currently, there is no interactive VR application on the market to be used to generate ultra-low latency traffic for the tests\footnote{Current available VR applications such as Youtube360 are not delay sensitive, because they simply send all 360-degree view (corresponding video frames) at once to the end-user. Therefore, they consume way more bandwidth}. Therefore, to evaluate BoDe, we use three different categories of applications:
\begin{enumerate}
\item We use available interactive applications with low latency requirements (namely Skype and Google Hangout).
\item We use video streaming applications (YouTube and AmazonPrime) and various ABR algorithms to emulate throughput behavior of future VR interactive applications which will likely be equipped with advanced mechanisms to change their sending rates (video quality) according to the link quality. 
\item We use Iperf3 as our source of traffic and use various state-of-the-art transport control designs to show how transparent is BoDe to various end-host protocols.
\end{enumerate}  

\begin{figure}[!t]
    \centering
    \begin{minipage}[b]{0.48\linewidth}
            \includegraphics[width=\textwidth,height=1.5in]{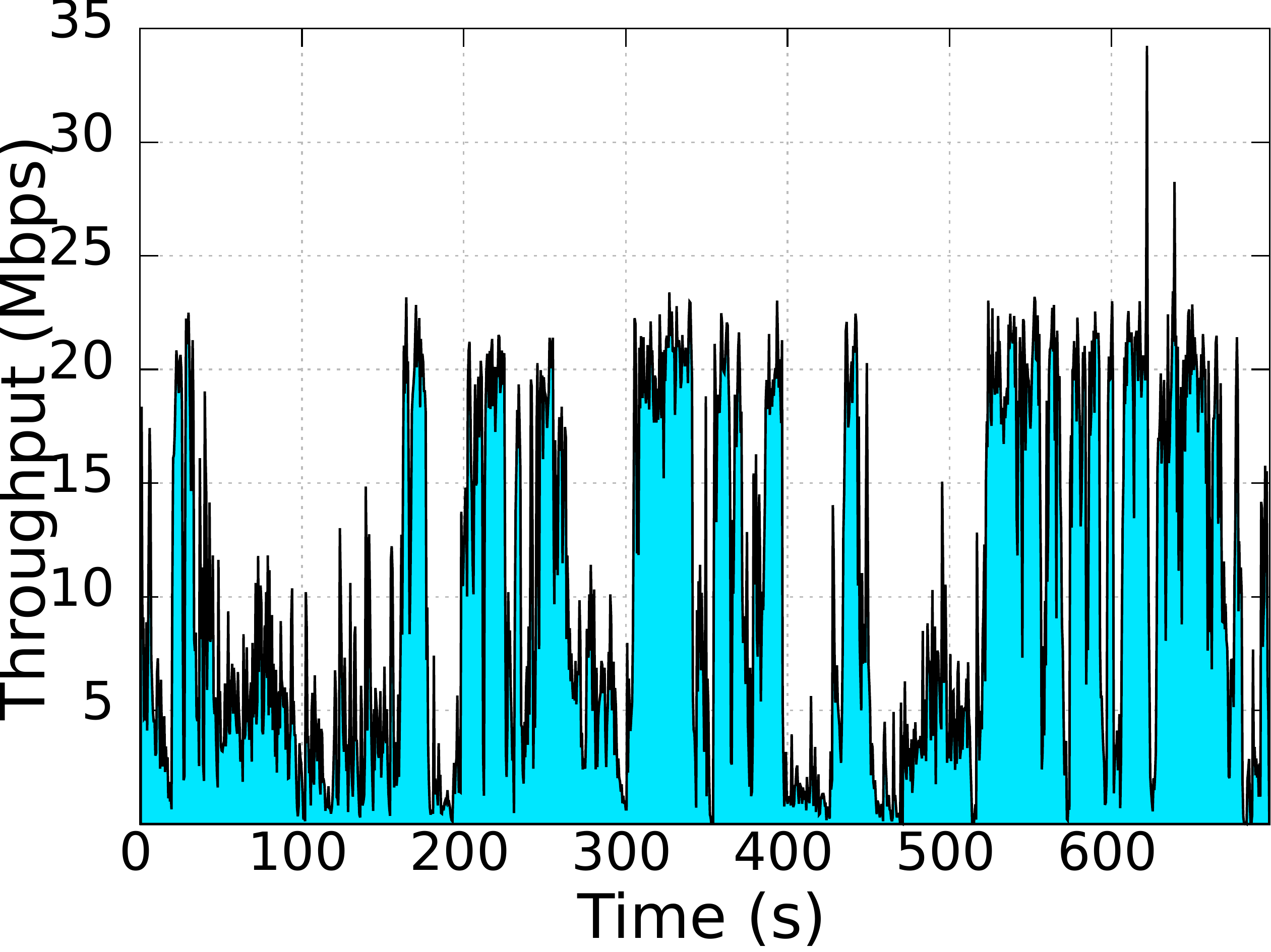}                 
            \subcaption{Trace gathered while riding a subway train in NYC}
            \label{fig_trace_times}
    \end{minipage}
    \hfill
    \begin{minipage}[b]{0.48\linewidth}
        \includegraphics[width=\textwidth,height=1.5in]{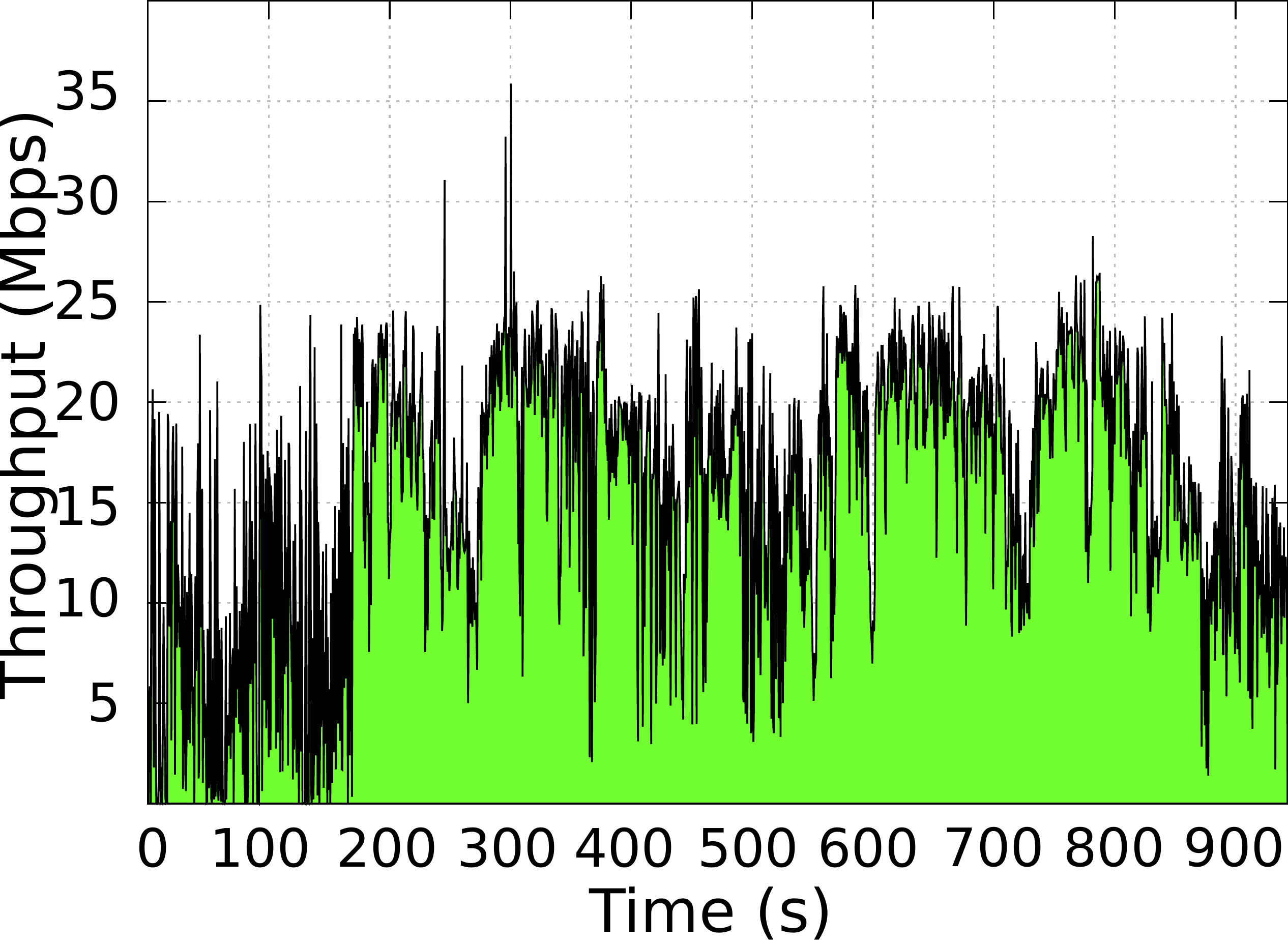}                         
        \subcaption{Trace gathered in Times Square}
        \label{fig_trace_times}
    \end{minipage}
%        \includegraphics[width=0.8\textwidth,height=1.2in]{abr-qoe-delay}
%        \subcaption{Youtube}
        \caption{Samples of cellular LTE traces~\cite{c2tcp} used in evaluations}
        \label{fig_traces}
\end{figure}

\textbf{Cellular Traces:} We use cellular traces collected in prior work~\cite{sprout,c2tcp} as our last mile cellular links. Variation of bandwidth on two of these traces is shown in Fig.~\ref{fig_traces}. Generally, on each trace, we run each test for 5 minutes. 

\textbf{Schemes Compared:} We compare performance of various AQM designs including FIFO TailDrop, FIFO HeadDrop, CoDel\cite{codel}, and PIE\cite{pie} to BoDe~\footnote{ABC~\cite{abc}'s prototype implementation provided by its authors uses a hard-coded simple server-client application. So, the prototype version of ABC cannot be used for existing applications such as Iperf3, Skype, Hangout, etc. Also, their current code hard-coded over UDP and cannot be used to send TCP traffic. Hence, we couldn't include ABC in our experiments where we use real-world unmodified applications}. 

\textbf{Performance Metrics:} We use 3 main performance metrics in this section: average delivery rate (in short, throughput), 99th percentile queuing delay, and power defined as $\frac{throughput}{99th -\%tile-delay}$. 

\textbf{Topology:} The topologies of all three testbeds are shown in Fig.~\ref{fig_topo}. The minimum RTT delay is set to 10ms to emulate an MEC like network where servers are at/close to the mobile 
edge.

\textbf{Schemes' Parameters:} Table\ref{table_tune} shows the parameters that we used for schemes in our evaluations\footnote{We tuned parameters of CoDel and PIE based on our separate tuning experiment runs and based on their authors' recommendations to get (on average) their best performance}. Skype/Hangout can tolerate 100ms delays, so for the interactive evaluations, we set $D=100ms$. For other tests, we set $D=20ms$. Although Youtube and AmazonPrime Applications and Iperf3 are throughput hungry and not delay sensitive (Youtube and AmazonPrime buffer the stream for a few seconds before playing it and Iperf3  always have packets to send and have potentially high throughput), we set $D=20ms$ when testing these applications to evaluate BoDe's performance in the presence of applications which either can adapt themselves to the changes of the quality (YouTube and AmazonPrime) or can adjust their sending rates using TCP algorithms.

\begin{table}[!h] \renewcommand{\arraystretch}{1.2} \caption{Schemes' Parameters} 
\label{table_tune} 
%\footnotesize
\centering
\small
\begin{tabular}{c|c}
\hline
Scheme & Parameters \\
\hline
%\rowcolor{LightBlue}
BoDe& \textit{BoundedDelay}=$D$\\
\hline
CoDel& min Delay=$D/2$, Interval=$5\times RTT$\\
\hline
PIE& \makecell{ Ref. Delay=$D$, $\alpha$=0.125,$\beta$=1.25} \\ %\\ $T_{update}$=30ms,max_burst=3ms}\\
\hline
Tail/HeadDrop& QueueSize=$1.5MB$\\
%\hline
%& QueueSize=$1.5MB$\\
\hline
\end{tabular}
\end{table}

\subsection{Interactive Applications}
\label{sec_inter}
\begin{figure*}[!t]
\centering
    \begin{minipage}[t]{0.32\linewidth}
    \includegraphics[width=0.95\textwidth,height=1.5in]{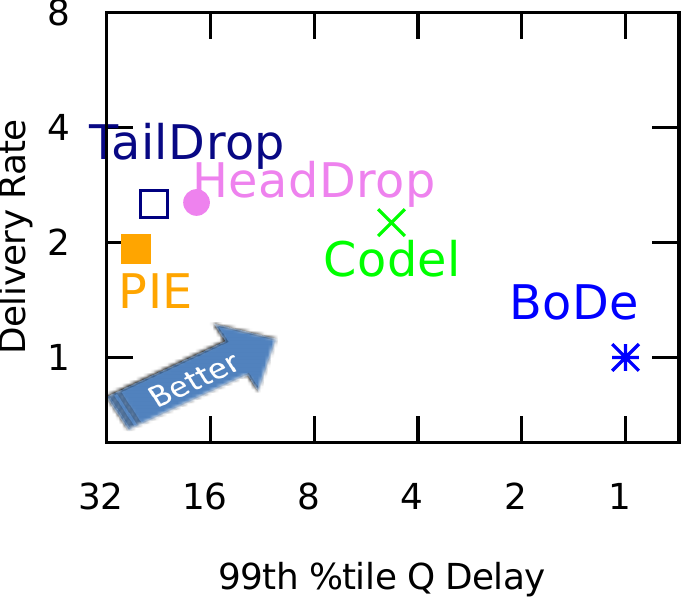}
    \subcaption{Skype}\label{fig_skype}
    \end{minipage}
    \hfill
    \begin{minipage}[t]{0.32\linewidth}
        \includegraphics[width=0.95\textwidth,height=1.5in]{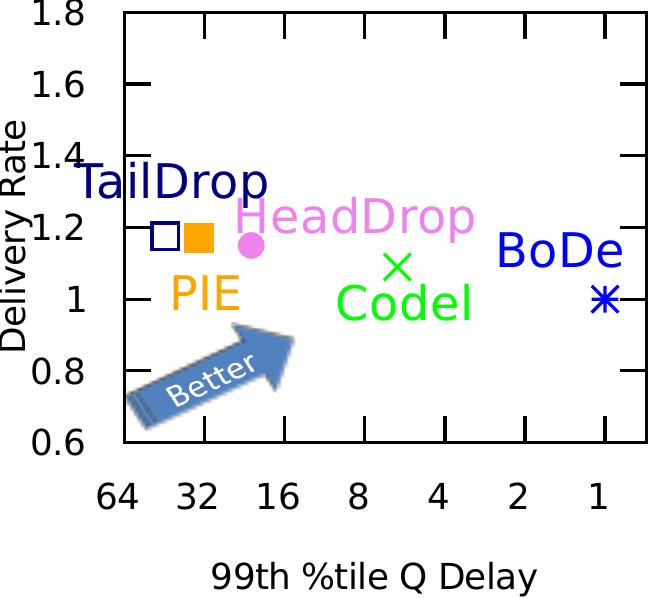}
        \subcaption{Hangout}
        \label{fig_hangout}
    \end{minipage}
    \hfill
    \begin{minipage}[t]{0.32\linewidth}
        \includegraphics[width=\textwidth,height=1.5in]{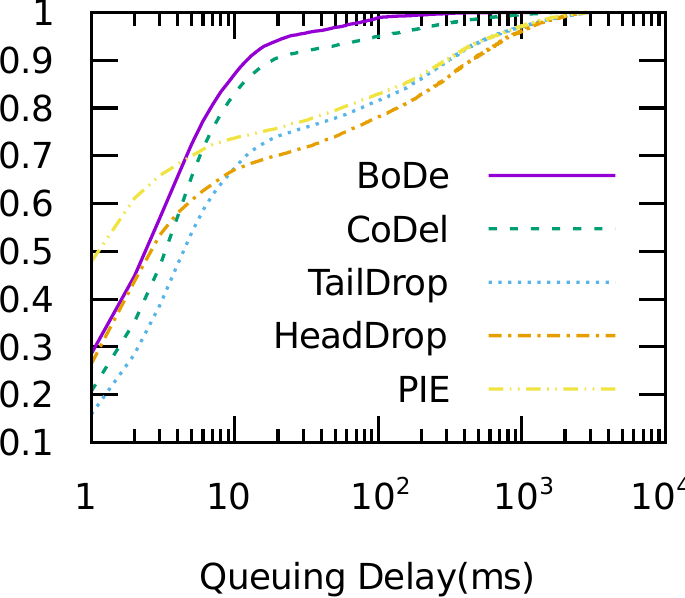}
        \subcaption{CDF of queuing delay for Skype traffic over T-Mobile Times Square cellular trace~\cite{c2tcp}}
        \label{fig_cdf-skype}
    \end{minipage}
    \caption{Normalized delay \& delivery rate of applications averaged over all traces and samples of CDF of queuing delay for specific applications and traces}
\end{figure*}

Here, we use Skype and Google Hangout as our interactive applications (The setup is shown in Fig.~\ref{fig_topo} (middle one)). To make sure that the only bottleneck is the network and not the quality of the source's webcam or receiver's display, we use 4K camera/4k display to generate/play the traffic.
For each cellular trace, we normalize the results of each scheme to the BoDe's results and average normalized results over all traces. Figures \ref{fig_skype} and \ref{fig_hangout} show the final averaged normalized results. BoDe achieves more than $4\times$ better queuing delay when compared to the 2nd best-performing scheme (CoDel), while it compromises throughput about $2\times$ (for Skype tests). To have a better understanding of delay performance, we report CDF of queuing delay of Skype traffic for one of the traces (Times-square-downlink~\cite{c2tcp}) in Fig.~\ref{fig_cdf-skype}. As Fig.~\ref{fig_cdf-skype} illustrates, BoDe bounds the overall queuing delay very well, while other schemes can cause delays up to a few seconds. It is worth mentioning that when the available capacity is very low, the transmission time of a packet increases, therefore queuing delay of that packet (consisting of sojourn time and transmission/service time) increases. BoDe only controls sojourn time of packets. Therefore, the overall queuing delay can be larger than \textit{BoundedDelay}. In addition, since BoDe always transmits the last 3 queued packets, occasionally, sojourn time of these packets could be larger than \textit{BoundedDelay}. However, even with these slightly larger delay sources, the overall performance compared to other schemes is still very well. For instance, considering Fig.~\ref{fig_cdf-skype}, the 99th percentile delay of BoDe is $107ms$ while this number is $620ms$ and $1704ms$ for CoDel and PIE respectively. 

\begin{figure*}[!t]
\centering
    \begin{minipage}[t]{0.32\linewidth}
    \includegraphics[width=0.95\textwidth,height=1.5in]{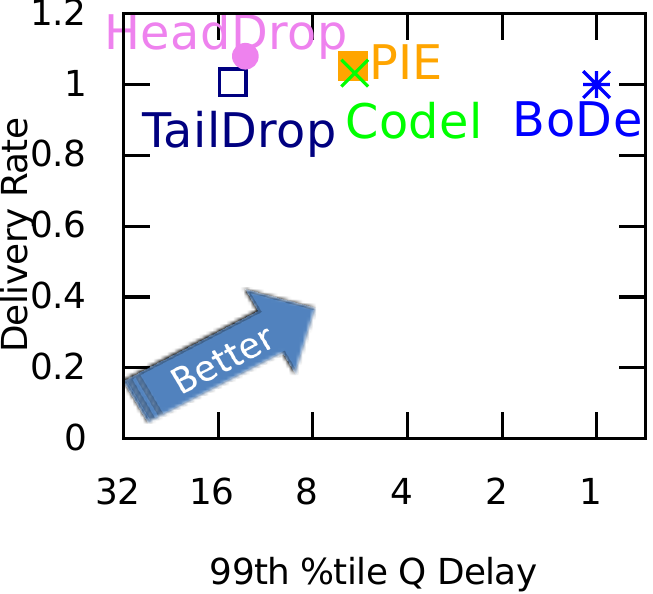}
    \subcaption{Youtube}
    \label{fig_youtube}
    \end{minipage}
    \hfill
    \begin{minipage}[t]{0.32\linewidth}
        \includegraphics[width=0.95\textwidth,height=1.5in]{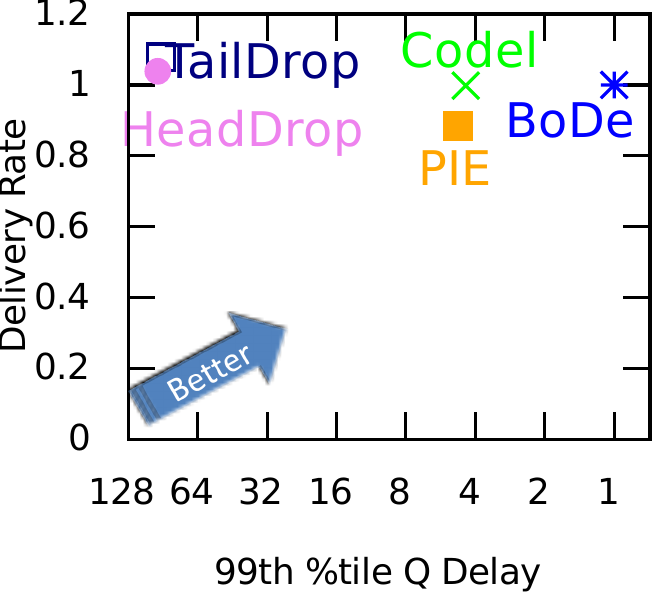}
        \subcaption{AmazonPrime}
        \label{fig_amazon}
    \end{minipage}
    \hfill
    \begin{minipage}[t]{0.32\linewidth}
        \includegraphics[width=\textwidth,height=1.5in]{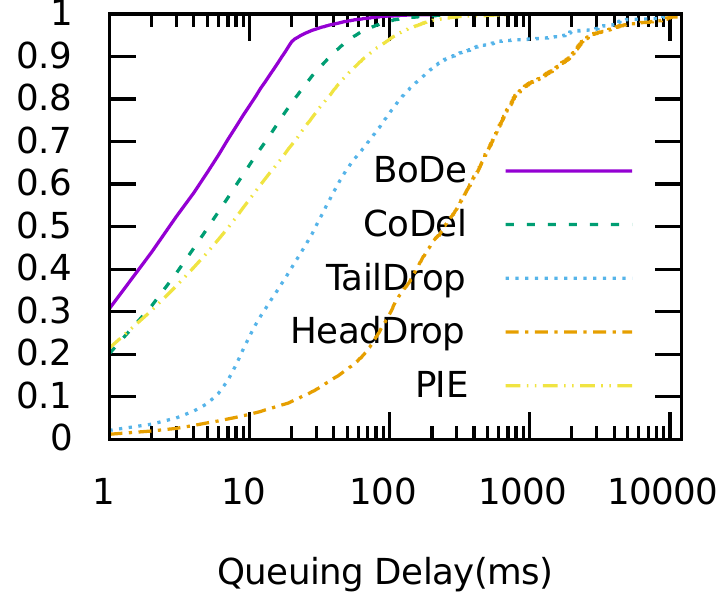}
        \subcaption{CDF of queuing delay for AmazonPrime over T-Mobile Subway cellular trace~\cite{c2tcp}}
        \label{fig_cdf-amazon}
    \end{minipage}
    \caption{Normalized delay \& delivery rate of applications averaged over all traces and samples of CDF of queuing delay for specific applications and traces}
\end{figure*}

\subsection{Video Streaming/ABR Applications}
Here, we use YouTube and AmazonPrime (as streaming applications) and various ABR algorithms including BB\cite{bb}, FESTIVE\cite{festive}, BOLA\cite{bola}, and robustMPC\cite{mpc}. Although having very large buffers at the receiver side makes these applications/algorithms delay-insensitive, our goal here is to show that for realistic variable bit rate traffic, by compromising a bit of throughput (by dropping packets properly), we can achieve significantly better delay responses.

For Amazon/YouTube tests, we use videos (with multiple available bit rates (up to 4k)) from Amazon/YouTube servers played on Firefox browser (The setup is shown in Fig.~\ref{fig_topo} (the right one)). 
For each trace, we normalize the results of each scheme to the BoDe's results and average normalized results over all traces. Figures ~\ref{fig_youtube} and ~\ref{fig_amazon} show the results. Also, CDF of queuing delay of AmazonPrime traffic for one of the traces (NYC-subway-downlink~\cite{c2tcp}) is shown in Fig.~\ref{fig_cdf-amazon}. BoDe outperforms other schemes. For instance, compared to the 2nd best-performing scheme (CoDel), it achieves more than $4\times$ better delay, while it compromises throughput less than $5\%$.

\begin{figure*}[!t]
    \centering
    \begin{minipage}[b]{0.48\linewidth}
            \includegraphics[width=.95\textwidth,height=1.5in]{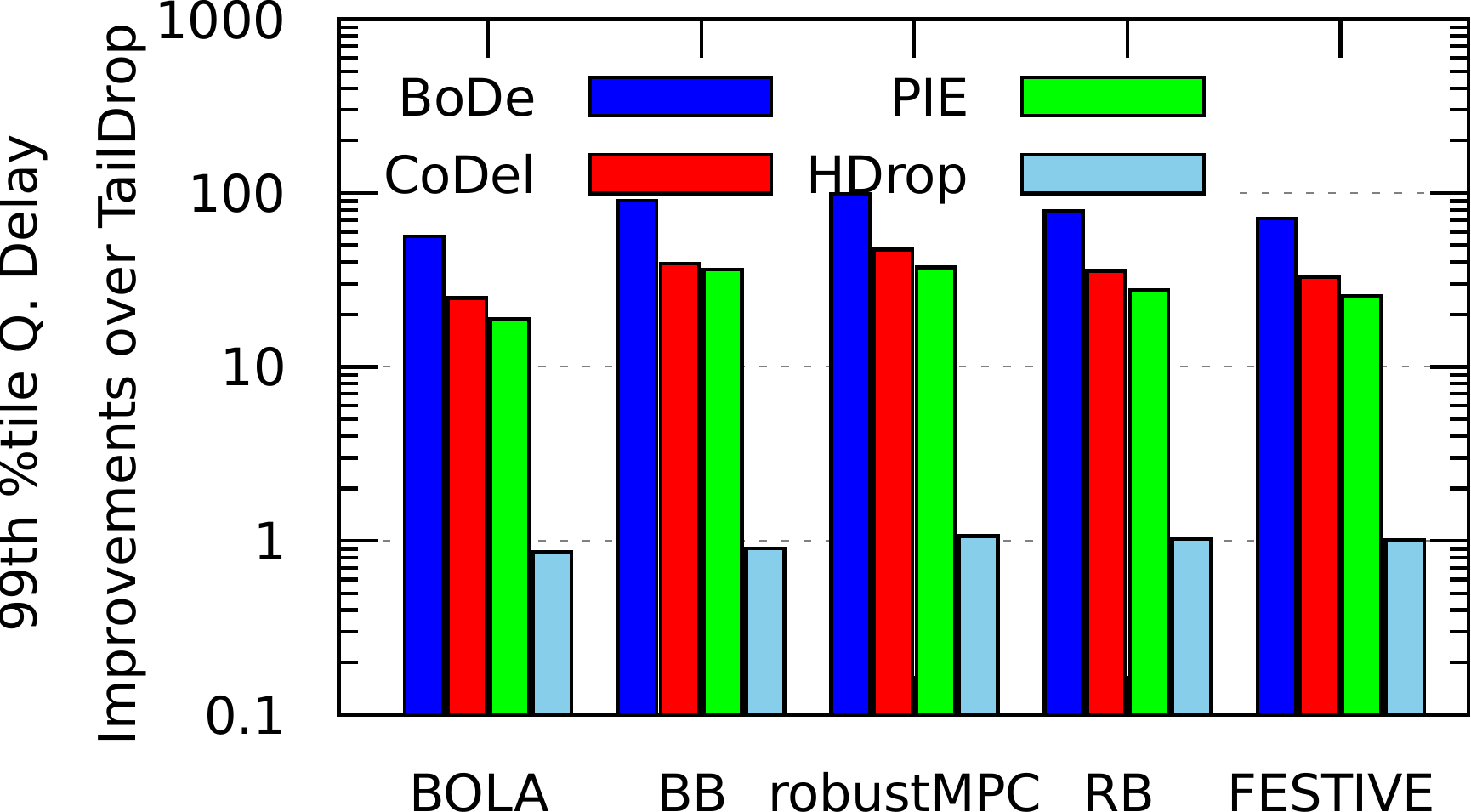}
    \end{minipage}
%    \hfill
    \begin{minipage}[b]{0.48\linewidth}
        \includegraphics[width=0.95\textwidth,height=1.5in]{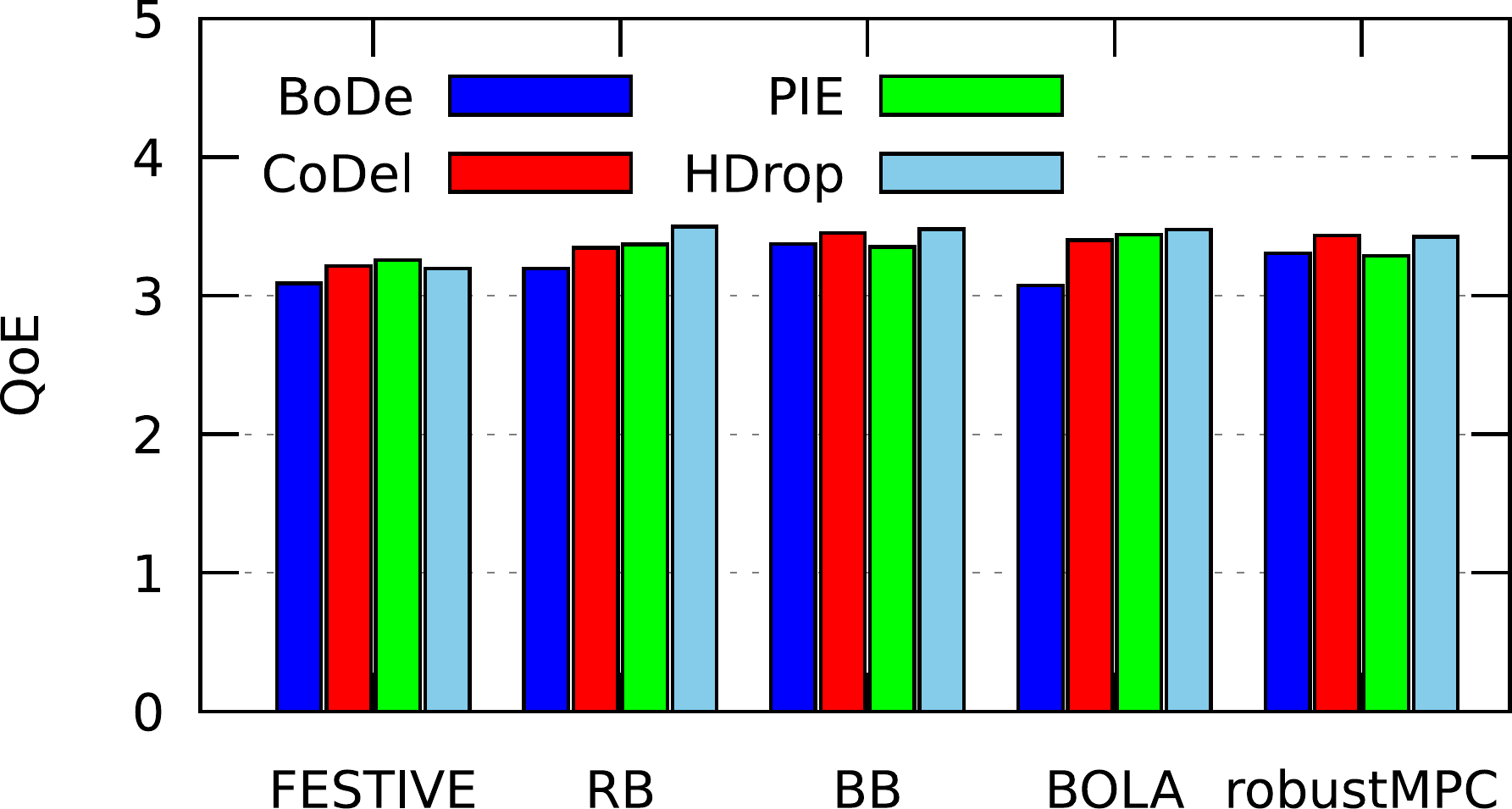}
    \end{minipage}
%        \includegraphics[width=0.8\textwidth,height=1.2in]{abr-qoe-delay}
%        \subcaption{Youtube}
        \caption{Ratio of 99th \% Q. delay improvements (left) \& QoE of various ABR alg. (right) averaged over all traces}
        \label{fig_abr}
\end{figure*}

For ABR tests, we use a modified version of dash.js supporting different ABR algorithms \cite{pens}. We use 48 video chunks with a total length of 193 seconds where each chunk has encoded in multiple bitrates to represent modes in \{240, 360, 480, 720, 1080, 1440\}p. In our setup, the client video player was a Firefox browser and the video server was an Apache server. We use QoE defined in \cite{mpc} as the user preference metric\footnote{Rebuffering penalty is set to 4.3.} and compare the 99th \%tile queuing delays improvements compared to the baseline TailDrop scheme (Fig.~\ref{fig_abr}). BoDe at least achieves $2\times$ lower 99th \%tile delay, while only compromising at most less than $0.1\times$ of QoE. 
\begin{figure*}[tp]
\centering
    \begin{minipage}[b]{0.48\linewidth}
    \includegraphics[width=0.95\textwidth,height=1.5in]{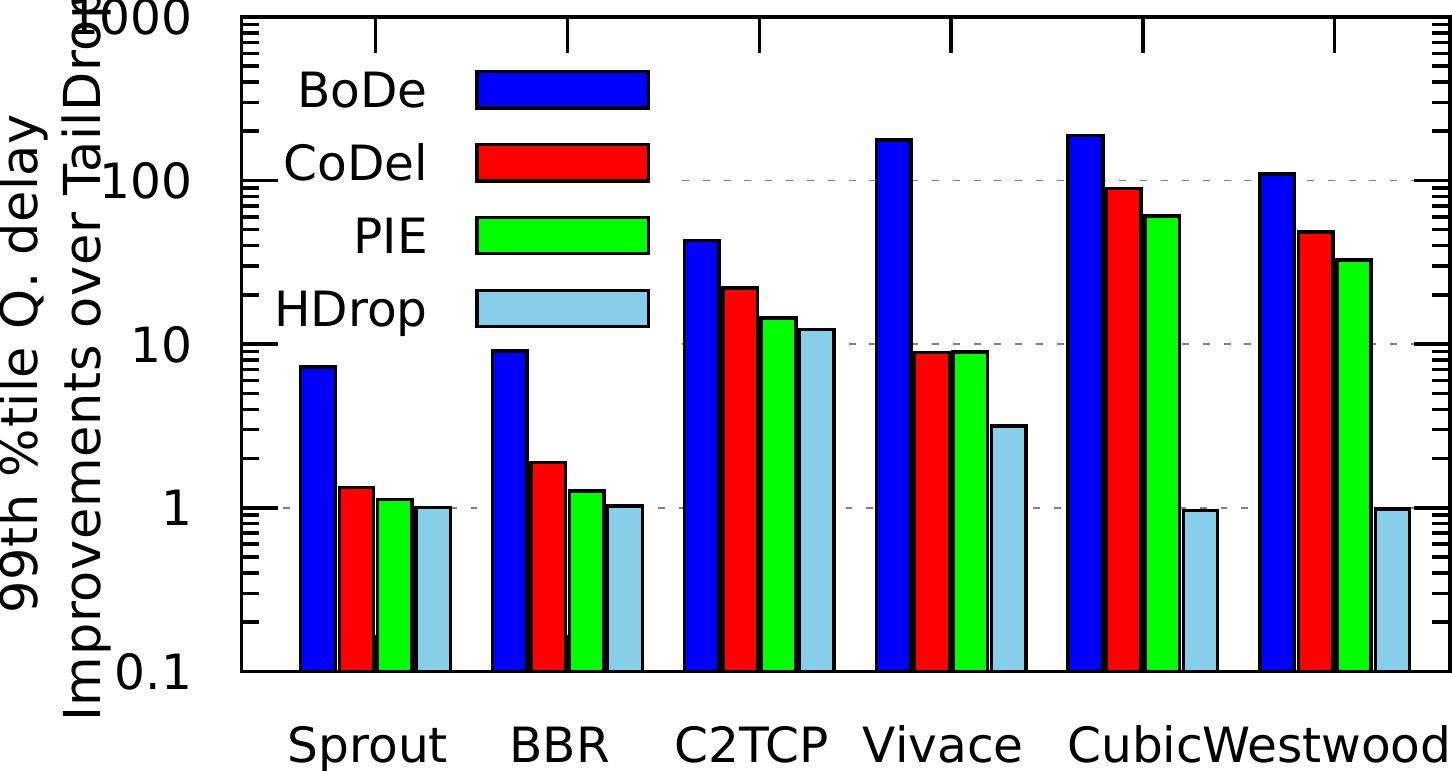}
    \end{minipage}
%    \hfill
    \begin{minipage}[b]{0.48\linewidth}
        \includegraphics[width=0.95\textwidth,height=1.5in]{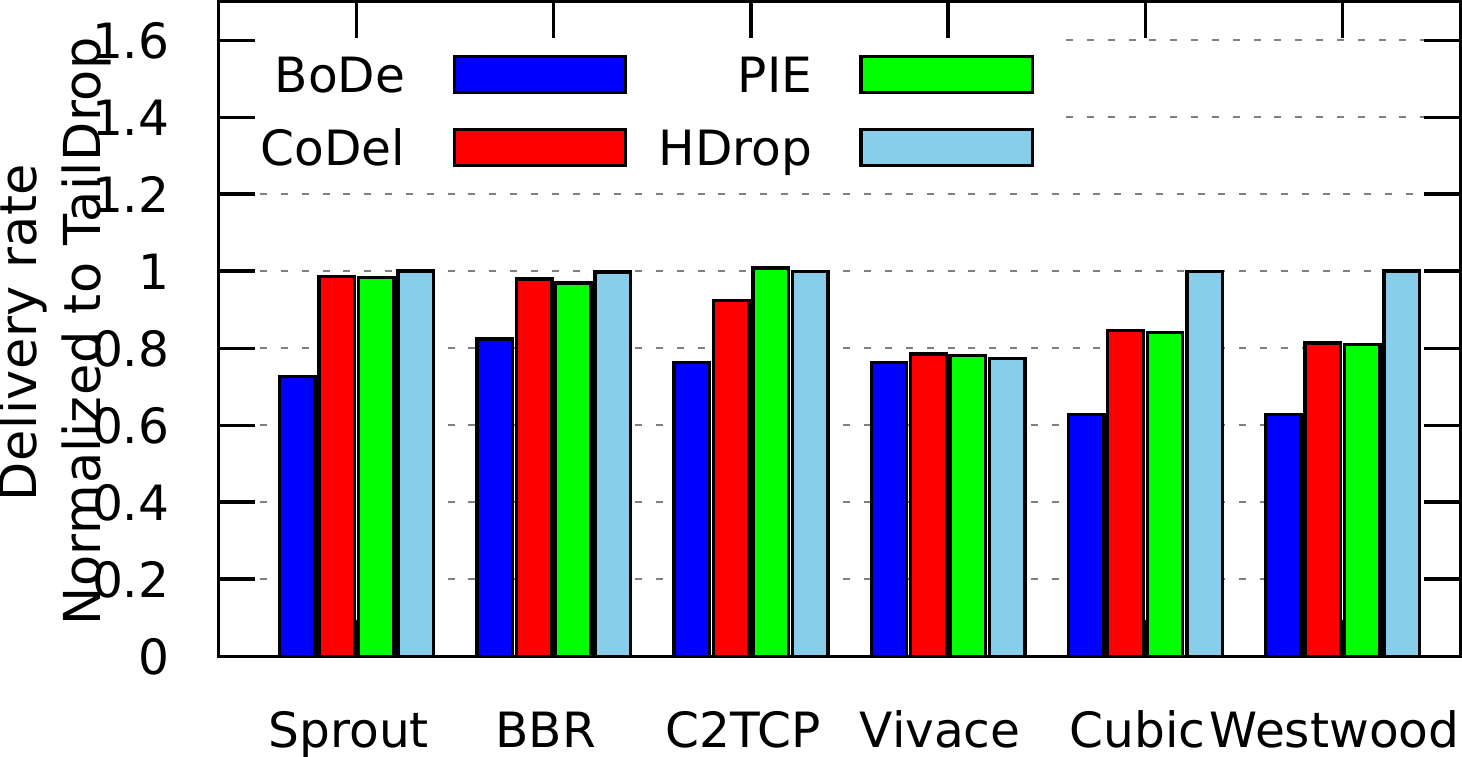}
%        \label{fig_tcp-thr}
    \end{minipage}
    \caption{The 99th \%tile queuing delay improvements (left) and the delivery rate (right) of various schemes normalized to FIFO-TailDrop performance and averaged over all traces}\label{fig_tcp}
\end{figure*}

\begin{figure}[tp]
\centering
       \includegraphics[width=.45\textwidth,height=1.5in]{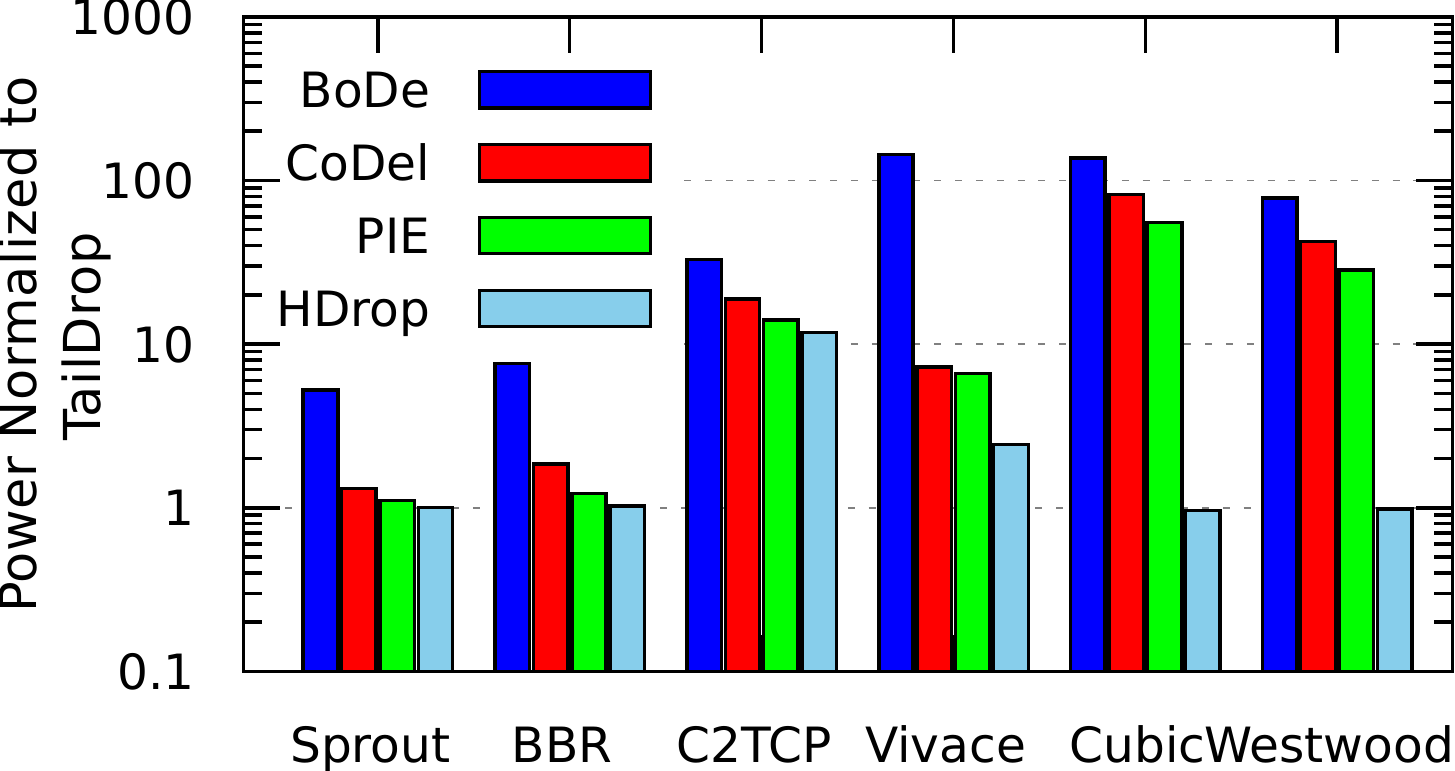}
    \caption{The power of various TCP schemes normalized to FIFO-TailDrop performance and averaged over all traces}\label{fig_tcp2}
\end{figure}

\subsection{Variety of Transport Control Schemes}
We use Google's BBR \cite{bbr}, PCC-Vivace~\cite{vivace}, C2TCP~\cite{c2tcp}, Sprout~\cite{sprout}, Cubic~\cite{cubic}, and Westwood~\cite{west} to evaluate BoDe's performance in the presence of different e2e TCP protocols (The setup is shown in Fig.~\ref{fig_topo} (left graph)). Fig.~\ref{fig_tcp} and Fig.~\ref{fig_tcp2} show the results of performance improvements normalized to TailDrop FIFO and averaged over all traces. As described in Section~\ref{sec_aqm-tcp}, the performance of CoDel and PIE depends on the TCP design. For instance, considering the delay, in the cases of BBR and Sprout, CoDel and PIE perform roughly similar to TailDrop scheme. However, BoDe outperforms all AQM schemes for all TCP scenarios dramatically. For example, BoDe can achieve about $20\times$ lower 99th percentile queuing delay compared to CoDel and PIE under PCC-Vivace. This great performance on delay comes with a small relative compromise in the throughput. This throughput compensation ranges from 3\% to 40\%. Even in the worst case, considering both delay and throughput, BoDe achieves at least about $2\times$ higher power compared to the best performing AQM approach. 

\begin{figure}[!t]
    \centering
    \begin{minipage}[b]{0.48\linewidth}
    \includegraphics[width=\textwidth,height=1.5in]{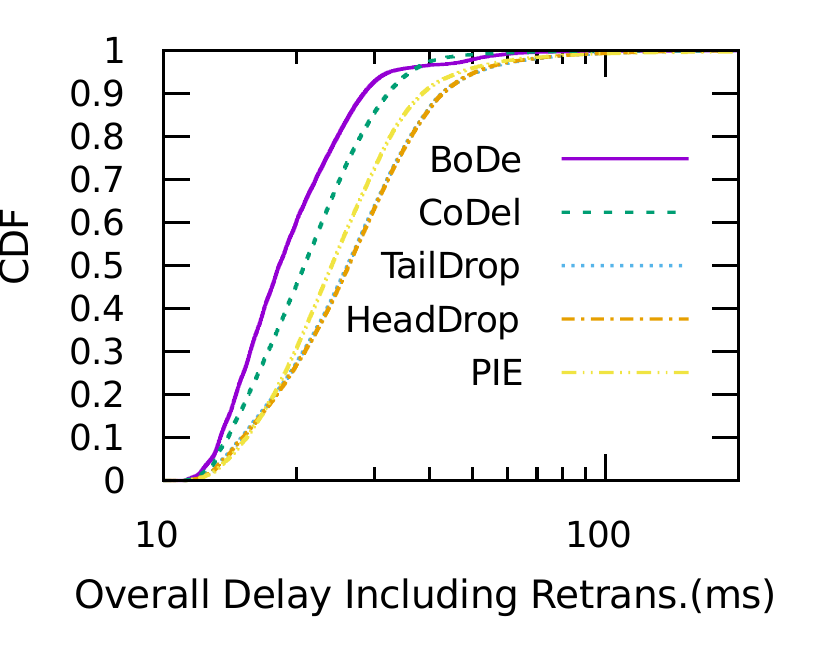}
    \end{minipage}
    \hfill    
    \begin{minipage}[b]{0.48\linewidth}
       \includegraphics[width=\textwidth,height=1.5in]{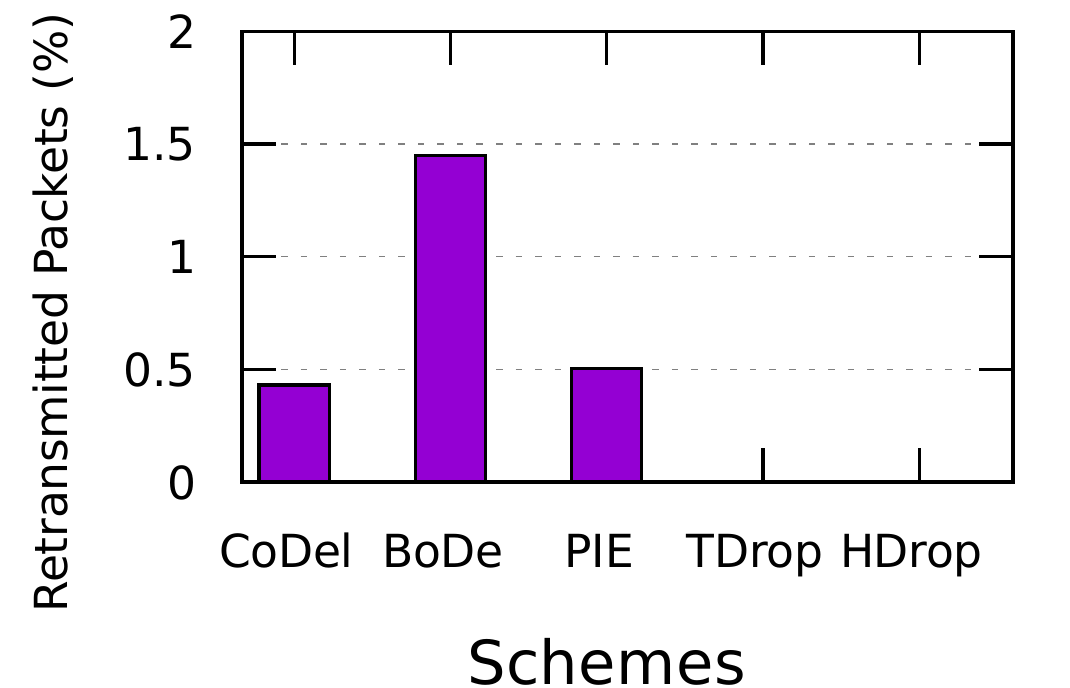}
    \end{minipage}
%    \hfill
%    \begin{minipage}[b]{0.28\linewidth}
%        \includegraphics[width=\textwidth,height=1.2in]{trace-att}
%    \end{minipage}
     \caption{CDF of overall Delay including retrans. time (left graph) and the ratio of retrans. pkts (right graph) using C2TCP}
        \label{fig_rtt}
\end{figure}

\subsubsection{Overall RTT including retransmissions} Although BoDe targets ultra-low latency interactive applications, in which packet retransmission is less relevant (as in video conferencing apps), its main strategy of reducing self-inflicted queuing delay can help to reduce e2e delays even when packet retransmissions are considered. 
Here, to show that, without loss of generality, we use AT\&T-LTE downlink trace~\cite{sprout}, Iperf3 as the application, and C2TCP~\cite{c2tcp} as the TCP scheme. The CDF of per packet e2e delay including the time spent for its probable retransmission(s) and percentage of retransmitted packets for various AQM designs are shown in Fig.~\ref{fig_rtt}. 
In a general network, retransmissions are not favorable, because of likely large intrinsic RTTs. However, the key insight here is that when servers are close to the UEs (as in MEC), the dominant part of the delay becomes queuing delay. So, keeping queuing delay very small (by actively and wisely dropping packets in the network (Fig.~\ref{fig_rtt} right graph)) benefits the delay responses of all packets including the retransmitted ones (Fig.~\ref{fig_rtt} left graph). 
%(Fig.~\ref{fig_rtt} left graph) 

\section{Discussion}
To start the discussion, we first, briefly describe the QoS specifications that are defined in 3GPP. In the latest 3GPP specifications, a set of QoS class identifiers (QCI) assigned to each bearer are defined. More specifically, 3GPP TS 123.203 specification~\cite{3gpp} identifies 4 parameters for each QCI: 1) resource type 2) priority level 3) packet delay budget (PDB) 4) packet error loss rate (PELR). Resource type identifies whether a bearer is of type of guaranteed bit rate (GBR) or non-GBR. Priority levels associated to each QCI identify that the wireless scheduler can preempt the traffic of a QCI with lower priority to serve the traffic of a QCI with higher priority (the lowest priority level number corresponds to the highest priority). In a nutshell, PDB is a soft upper bound for the time that a packet may be delayed between the UE and the radio interface\footnote{More specifically, PDB is the delay between the UE and the PCEF (Policy and Charging Enforcement Function unit).}. The purpose of the PDB is to support the configuration of wireless scheduling (e.g. the setting of scheduling priority weights) and link layer functions (e.g. the mac layer Hybrid automatic repeat request (HARQ) target operating points). PELR defines an upper bound for a rate of non-congestion related packet losses. The purpose of the PELR is to allow for appropriate link layer protocol configurations (e.g. HARQ in E-UTRAN). 

\subsection{Does the notion of QoS defined in 3GPP specifications already resolve the issue?}
The answer is no, it does not. The 3GPP TS 123.203 defines a set of services not how to achieve them. For instance, it mentions ``The discarding (dropping) of packets is expected to be controlled by a queue management function, e.g. based on pre-configured dropping thresholds.''\footnote{The 3GPP TS 123.203: Page 51, Note 1~\cite{3gpp}.} As discussed in section~\ref{sec_intro}, one of the key problems is the choice of having fixed and pre-configured set of queue size (drop) thresholds to drop the packets. That is exactly where BoDe AQM scheme which does not require any parameter tuning comes to the play.

\subsection{If certain PDB and PELR are specified, doesn't the wireless scheduler try to get the packets out within that PDB? If yes, why do we still need a modern AQM scheme?'}
The answer is that the task of the wireless scheduler and the task of AQM scheme are orthogonal and none of them can replace the other. The key goal of defining PDB in 3GPP TS 123.203 is to have a criterion to break the tie among traffic coming from different users with the same QCI priority level and decide which user's traffic should be scheduled first. At the best case where all users' traffic is shaped nicely at the end-hosts and cellular access link bandwidths are high enough to support all incoming traffic, a sensible scheduler that respects the bandwidth and delay constraints will be enough to satisfy delay demands of the traffic. However, in practice where wireless link bandwidth oscillates a lot, the traffic coming from the end-hosts are not necessarily shaped nicely, and the all incoming traffic cannot be served in a timely manner, even a perfect scheduler cannot satisfy the delay/throughput demands of the traffic. A simple scenario where the scheduler cannot work is the example shown in Fig.~\ref{fig_skype-step}.

Another important point is the subtle difference between PELR definition and possible drop rate of packets in BoDe. PELR, as defined in 3GPP TS 123.203, identifies a possible drop rate of packets not caused by congestion. In other words, PELR is only dedicated to the possible drop rate of packets in the wireless channel (due to the fading, bad quality of channel, etc.) not drop rate of packets in the queues.  

\subsection{Coexistence of Multiple Classes of Traffic}
\label{sec_multi}
Supporting various applications with different delay constraints is not feasible when all these applications are going to be queued in the same buffer. A simple example is when a bandwidth hungry application and a delay sensitive application are placed in the same buffer. In this scenario, the bandwidth-hungry application will fill up the buffer and make the delay-sensitive application experience a large queuing delay which dramatically impacts the overall delay of it. The notion of having different QCIs assigned to each bearer as defined in 3GPP TS 123.203 helps to resolve the coexistence issue of multiple classes of traffic. 

A practical solution is to combine DiffServ~\cite{DiffServ} architecture, 3GPP TS 123.203 definition of traffic classes, and BoDe. DiffServ is a simple strict priority queue mechanism which is already available in the commodity switches to serve various classes of applications. More specifically, first, separate DiffServ classes corresponding to separate QCIs will be defined. For instance, without loss of generality, let's assume that there are 3 classes defined as follow.

\begin{itemize}
\item Class \#1: Including applications with network delay requirement of less than 50ms 
\item Class \#2: Including applications with network delay requirement of less than 100ms 
\item Class \#3: Including applications without any delay requirement
\end{itemize}

Then, using differentiated services code point (DSCP) in the IP header, packets belonging to every class will be tagged by the servers. Using the tagged priorities, packets will be placed in 3 separate queues in the network. Later, queues corresponding to different classes will be served in the order of their priority level (Class \#1 followed by Class \#2 followed by Class \#3). Queues corresponding to Class \#1 and Class \#2 traffic will be managed using separate BoDe algorithms with \textit{BoundedDelay}s equal to 50ms and 100ms, respectively. However, since Class \#3 traffic is not delay-sensitive, BoDe will not be used as AQM in queue \#3. Instead, Class \#3 traffic can be simply managed using conventional throughput-oriented AQM techniques such as a simple TailDrop queue.

Now, we conduct an experiment to show the benefits of using BoDe in the scenario of coexistence of multiple classes of traffic. We compare the performance of the DiffServ architecture using TailDrop FIFO queues (we call this scheme: DiffServ+FIFO) and DiffServ architecture using BoDe and TailDrop queues (we call this scheme: DiffServ+BoDe). In particular, we consider the 3 classes described above and use BoDe only for the two highest priority ones. We use UDP traffic sessions with average rates of 2Mbps and 4Mbps as Class \#1 and Class \#2 traffic respectively. For Class \#3, we use a TCP Cubic flow representing throughput-oriented file-download traffic. 
\begin{figure*}[!t]
\centering
\begin{minipage}[b]{\linewidth}
    \begin{minipage}[b]{0.48\linewidth}
    \includegraphics[width=0.99\textwidth,height=2.7in]{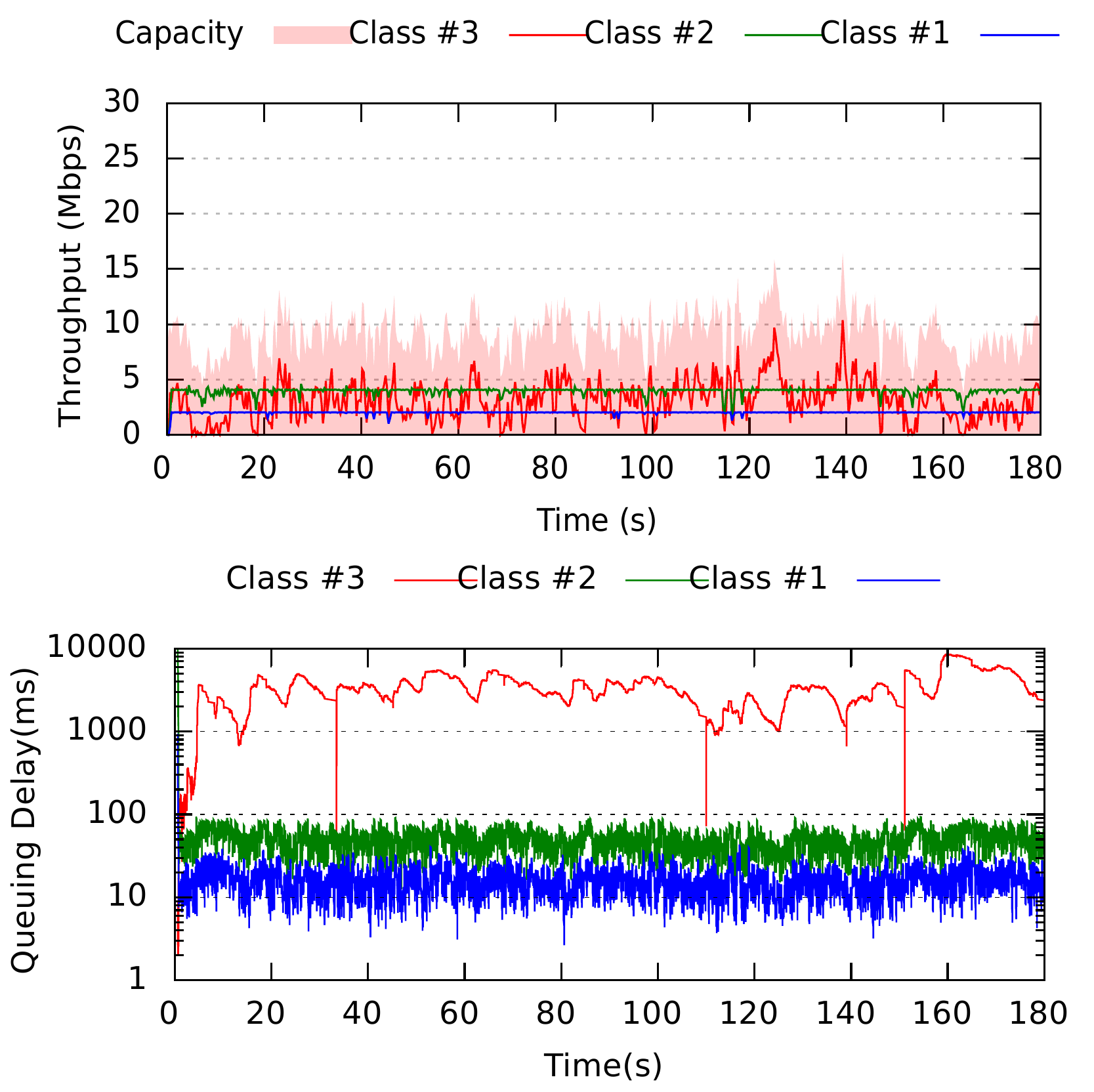}
    \subcaption{DiffServ+BoDe}
    \label{fig_bode_times}
    \end{minipage}
    \hfill
    \begin{minipage}[b]{0.48\linewidth}
        \includegraphics[width=0.99\textwidth,height=2.7in]{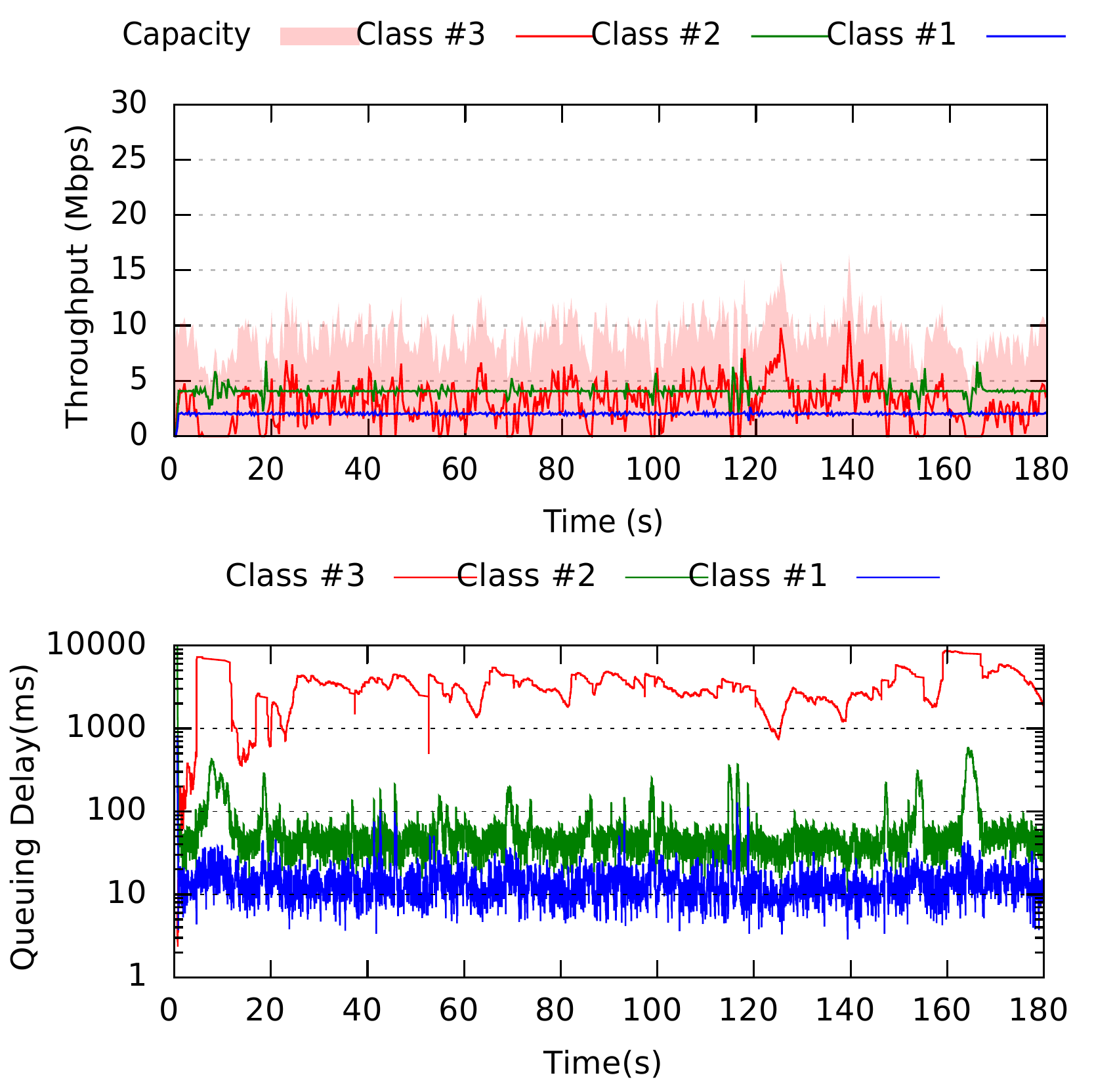}
        \subcaption{DiffServ+FIFO}
        \label{fig_fifo_times}
    \end{minipage}
    \caption{Queuing delay (bottom graphs) and delivery rate (top graphs) of different traffic classes for T-Mobile stationary trace using DiffServ architecture with different AQM schemes}
     \label{fig_station}
\end{minipage}
%\end{figure*}
%\begin{figure*}[!t]
%\centering
\begin{minipage}[b]{\linewidth}
    \begin{minipage}[b]{0.48\linewidth}
    \includegraphics[width=0.99\textwidth,height=2.7in]{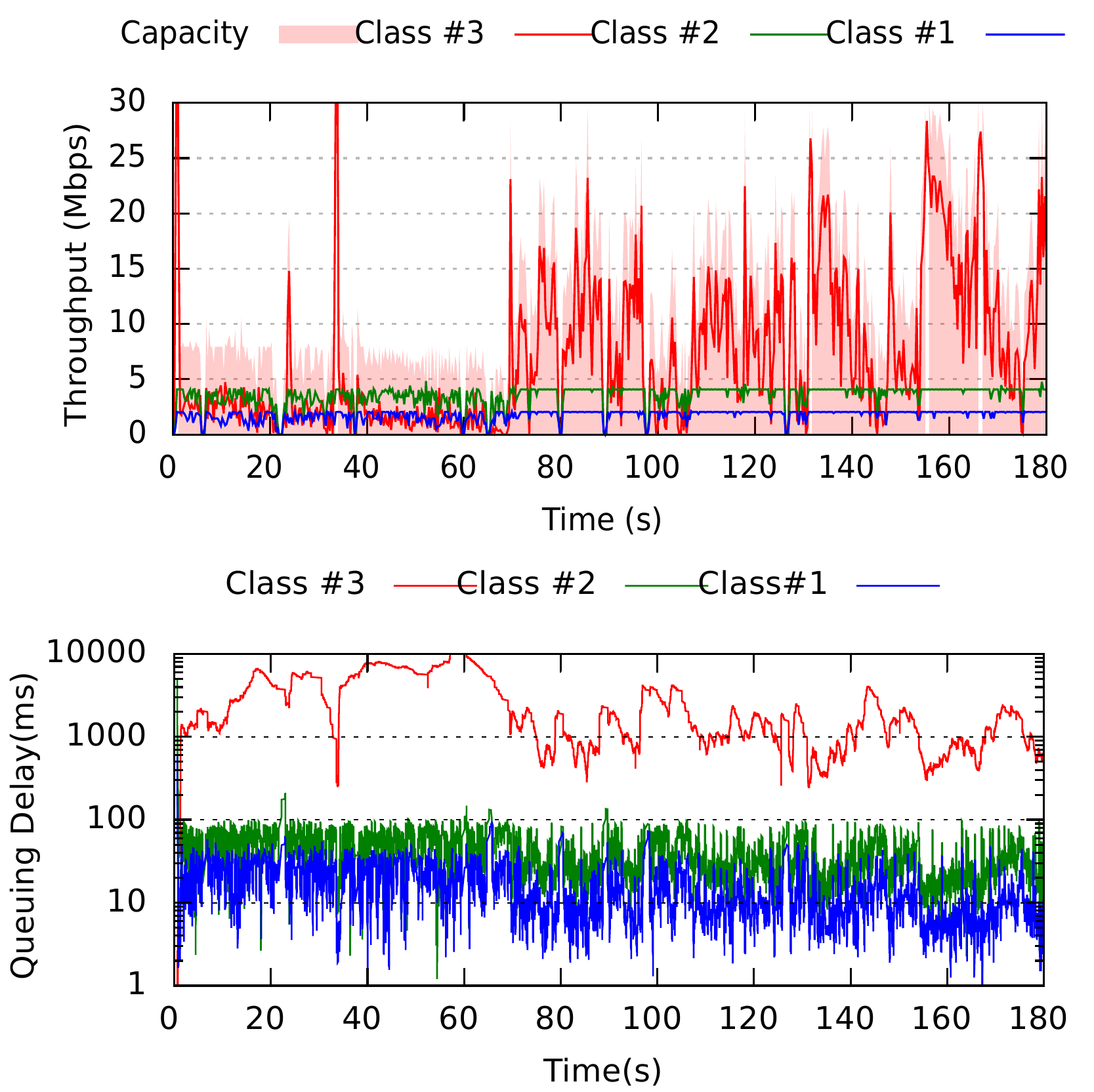}
    \subcaption{DiffServ+BoDe}
    \label{fig_bode_driving}
    \end{minipage}
    \hfill
    \begin{minipage}[b]{0.48\linewidth}
        \includegraphics[width=0.99\textwidth,height=2.7in]{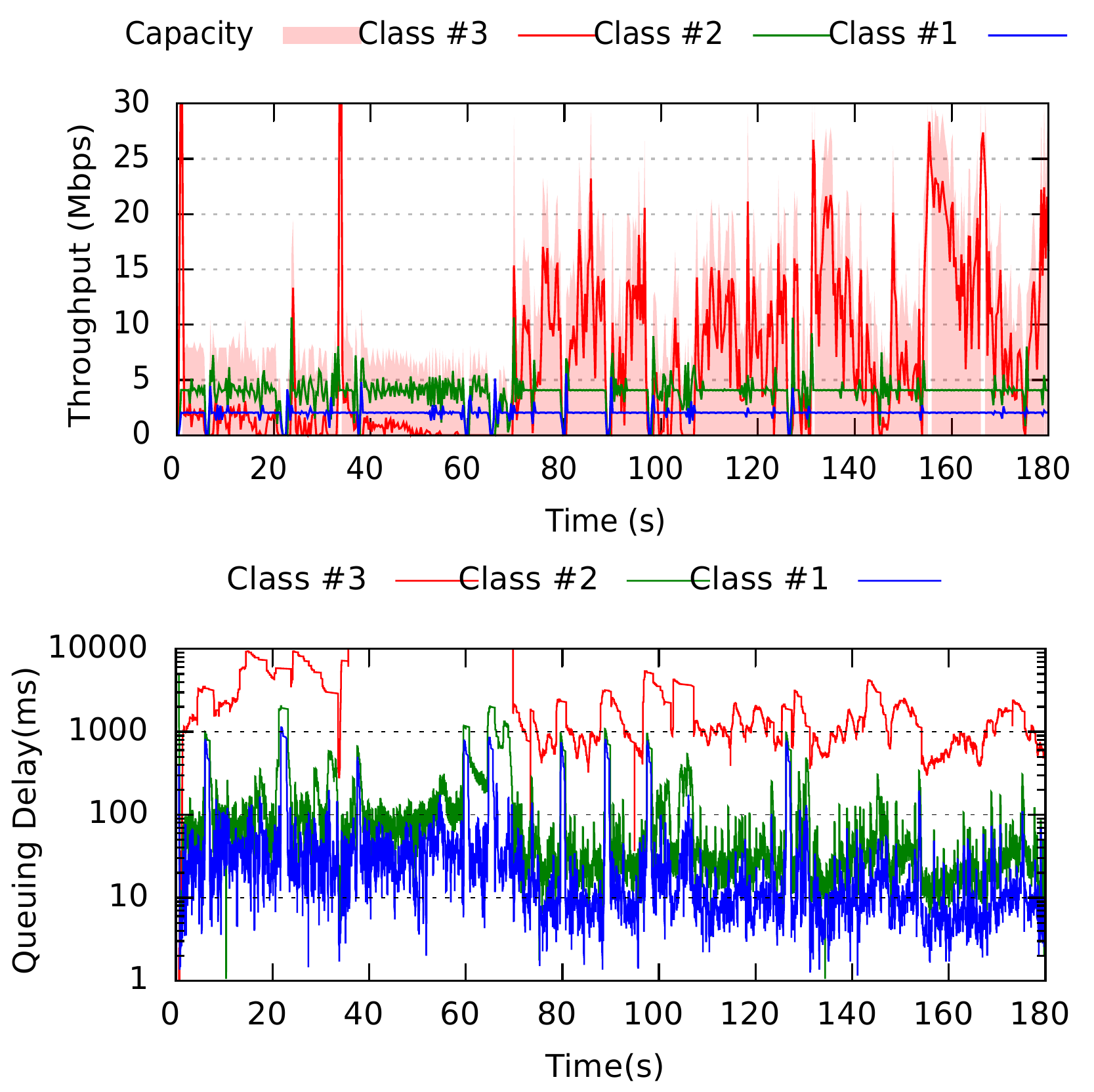}
        \subcaption{DiffServ+FIFO}
        \label{fig_fifo_driving}
    \end{minipage}
    \caption{Queuing delay (bottom graphs) and delivery rate (top graphs) of different traffic classes for T-Mobile driving trace using DiffServ architecture with different AQM schemes}
    \label{fig_driving}
\end{minipage}
\end{figure*}

Drop thresholds (buffer sizes) will impact the performance of TailDrop queues. So, for DiffServ+FIFO scheme, first, we carefully tune these thresholds using an arbitrary cellular network condition (an arbitrary cellular trace, e.g. T-Mobile stationary trace) to achieve delay/throughput performances similar to the performances gained by DiffServ+BoDe for the same network condition (Fig.~\ref{fig_station}). We use these setting throughout this experiment.

Now, we use another cellular trace (T-Mobile driving trace) representing another network condition for the same network provider and compare the performance of two DiffServ+BoDe and DiffServ+FIFO. As Fig.~\ref{fig_driving} illustrates, DiffServ+BoDe achieves desired bounded delay for both Class \#1 and \#2 traffic while DiffServ+FIFO fails to do that. The key reason is the fact that cellular networks intrinsically experience highly variable access link bandwidths which makes them very difficult to be tuned for various network conditions. 

\subsection{Is BoDe a solution for all applications?}
BoDe is a design targeting the delay-sensitive applications (similar to Class \#1 and Class \#2 traffic mentioned in section~\ref{sec_multi}) and not the traditional delay-insensitive applications (similar to Class \#3 mentioned in section~\ref{sec_multi}). The key reason is that all packets of a delay-insensitive application have the same level of importance and \textit{all} of them need to be received to have a successful transmission of data. The use of BoDe for these kinds of applications will not increase their performance which solely depends on their achieved throughput. 

However, the key property of delay-sensitive applications is that more recent packets are strictly more important than the older ones. In other words, the importance of packets depends on the time when they are generated. A simple example is when a user is using a video conferencing application in which the higher level of interactivity (receiving packets in a timely manner) has always higher priority compared to having a higher quality video. Therefore, for delay-sensitive applications trading a bit of throughput (by dropping packets wisely) to achieve lower delay is desired. 

In sum, BoDe is not a solution for all types of applications. BoDe is not designed to boost the throughput of delay-insensitive applications (such as web applications, streaming applications, etc.). It is designed to boost the delay performance of delay-sensitive applications (such as interactive applications (e.g. video conferencing, VR/AR, etc.)). Therefore, in practical mixed-traffic scenarios, conventional throughput-oriented AQM designs will be used for the queues serving delay-insensitive traffic (e.g. a simple TailDrop approach used for Class \#3 in section~\ref{sec_multi}), while designs such as BoDe will be applied to the queues serving delay-sensitive traffic. 

\subsection{Impact of BoDe on Buffer Size}
BoDe postpones the drop decision from the time of enqueuing the packet to the time of dequeuing it. So, it requires enough buffer size to absorb the packets. To see the impact of BoDe on the buffer size,  we assume the following: 
\begin{enumerate}
\item Drop of a packet does not take time (compared to the inter-arrival of the packets, drop time is negligible). 
\item If a packet is dropped, the next packet in the queue will be processed immediately. 
\item To have a stable system, the arrival rate of packets is less than the service rate of them (long-term cellular access link bandwidth). 
\end{enumerate}  

Therefore, using the Little's law theorem, the long-term maximum length of the queue will be the long-term maximum capacity of the cellular access link multiplied by the maximum delay that packets can experience in the queue ($BoundedDelay$). For instance, for a cellular access link with maximum access link of 96Mbps, the packet size of 1.5KB, and a traffic class with $BoundedDelay=10ms$, the buffer-size needed to absorb this class of traffic will be $\frac{96\times10^6}{1.5\times10^3\times8}\times0.01=80pkts=0.12MB$ which is very smaller than the large per-user buffers used at base stations~\cite{sprout}.

Moreover, the use of shared-memory structure at eNodeB for various users (similar to the commercial switches in which a pool of memory is shared among different inputs/outputs) alleviates the buffer-size usage, because the probability of having all users sending at the maximum rate simultaneously is very low (wireless scheduler at eNode schedules/limits the access of each user to the cellular channel periodically). 

\subsection{What If Buffer Size Is Way Smaller Than What BoDe Requires?}
Generally, choosing very small queue size leads to very low queuing delay, while it dramatically reduces the throughput of applications (due to the excessive inevitable drop of packets). Hence, if for any reason, buffer size is smaller than what BoDe requires to provide a certain $BoundedDelay$ for the queue, then the queuing delay will be even lower than the target $BoundedDelay$! In other words, with the very small buffer size, BoDe behaves similar to a TailDrop FIFO queue with a lower queuing delay than    $BoundedDelay$ value!

\subsection{What If Buffer Size Is Way Larger Than What BoDe Requires?}
A very important feature of BoDe is that the increase in the buffer size will not impact its performance. In a normal TailDrop FIFO queue with a large capacity, keeping more packets in the queue increases the waiting time of incoming packets and causes the well-known bufferbloat issue. However, in a BoDe queue, the packets that are waited for a long time in the queue will be dropped before being served. This will avoid the propagation of large waiting times in the queue which prevents the bufferbloat issue. 

\section{Conclusion}
In this paper, our key statement is that satisfying the ultra-low latency nature of interactive and real-time applications in a highly dynamic network such as the cellular network is only achievable by using proper AQM design in the network. No fully e2e scheme alone can meet ultra-low latency requirements and drop of packets in cellular networks is an inevitable important part of the solution. Based on that, we presented BoDe, an extremely simple yet powerful AQM design for current and future cellular networks to support ultra-low latency applications. 
%We showed that although BoDe is simple and stands on straightforward principles, it boosts the overall performance of the applications and dramatically outperforms the current state-of-the-art AQM schemes. 
We hope that the great delay performance of BoDe encourages cellular network providers to use delay-centric AQM designs such as BoDe to boost the user experience. 
%Also, we believe that achieving good performance does not necessarily come from complex drop rate calculation algorithms or complicated queue management techniques. We showed that although BoDe is simple and stands on straightforward principles, it boosts the overall performance of the applications and dramatically outperforms the current state-of-the-art AQM schemes.
%\vfill
	
%\bibliography{ref}
\bibliographystyle{IEEEtran}
%\bibliography{IEEEabrv,ref}

\begin{thebibliography}{10}
\providecommand{\url}[1]{#1}
\csname url@samestyle\endcsname
\providecommand{\newblock}{\relax}
\providecommand{\bibinfo}[2]{#2}
\providecommand{\BIBentrySTDinterwordspacing}{\spaceskip=0pt\relax}
\providecommand{\BIBentryALTinterwordstretchfactor}{4}
\providecommand{\BIBentryALTinterwordspacing}{\spaceskip=\fontdimen2\font plus
\BIBentryALTinterwordstretchfactor\fontdimen3\font minus
  \fontdimen4\font\relax}
\providecommand{\BIBforeignlanguage}[2]{{%
\expandafter\ifx\csname l@#1\endcsname\relax
\typeout{** WARNING: IEEEtran.bst: No hyphenation pattern has been}%
\typeout{** loaded for the language `#1'. Using the pattern for}%
\typeout{** the default language instead.}%
\else
\language=\csname l@#1\endcsname
\fi
#2}}
\providecommand{\BIBdecl}{\relax}
\BIBdecl

\bibitem{c2tcp}
\BIBentryALTinterwordspacing
S.~Abbasloo, Y.~Xu, and H.~J. Chao, ``C2tcp: A flexible cellular tcp to meet
  stringent delay requirements,'' \emph{IEEE Journal on Selected Areas in
  Communications}, 2019. [Online]. Available:
  \url{https://doi.org/10.1109/JSAC.2019.2898758}
\BIBentrySTDinterwordspacing

\bibitem{c2tcp2}
S.~Abbasloo, T.~Li, Y.~Xu, and H.~J. Chao, ``Cellular controlled delay {TCP}
  ({C2TCP}),'' in \emph{2018 IFIP Networking Conference (IFIP Networking) and
  Workshops}, 2018, pp. 118--126.

\bibitem{sprout}
K.~Winstein, A.~Sivaraman, H.~Balakrishnan \emph{et~al.}, ``Stochastic
  forecasts achieve high throughput and low delay over cellular networks.'' in
  \emph{NSDI}, 2013, pp. 459--471.

\bibitem{vivace}
M.~Dong, T.~Meng, D.~Zarchy, E.~Arslan, Y.~Gilad, B.~Godfrey, and M.~Schapira,
  ``$\{$PCC$\}$ vivace: Online-learning congestion control,'' in \emph{15th
  $\{$USENIX$\}$ Symposium on Networked Systems Design and Implementation
  ($\{$NSDI$\}$ 18)}, 2018, pp. 343--356.

\bibitem{pie}
R.~Pan, P.~Natarajan, C.~Piglione, M.~S. Prabhu, V.~Subramanian, F.~Baker, and
  B.~VerSteeg, ``Pie: A lightweight control scheme to address the bufferbloat
  problem,'' in \emph{High Performance Switching and Routing (HPSR), 2013 IEEE
  14th International Conference on}.\hskip 1em plus 0.5em minus 0.4em\relax
  IEEE, 2013, pp. 148--155.

\bibitem{codel}
K.~Nichols and V.~Jacobson, ``Controlling queue delay,'' \emph{Communications
  of the ACM}, vol.~55, no.~7, pp. 42--50, 2012.

\bibitem{cubic}
S.~Ha, I.~Rhee, and L.~Xu, ``Cubic: a new tcp-friendly high-speed tcp
  variant,'' \emph{ACM SIGOPS Operating Systems Review}, vol.~42, no.~5, pp.
  64--74, 2008.

\bibitem{tahoa}
V.~Jacobson, ``Congestion avoidance and control,'' in \emph{ACM SIGCOMM CCR},
  vol.~18, no.~4.\hskip 1em plus 0.5em minus 0.4em\relax ACM, 1988, pp.
  314--329.

\bibitem{newreno}
T.~Henderson, S.~Floyd, A.~Gurtov, and Y.~Nishida, ``The newreno modification
  to tcp's fast recovery algorithm,'' Tech. Rep., 2012.

\bibitem{bic}
L.~Xu, K.~Harfoush, and I.~Rhee, ``Binary increase congestion control (bic) for
  fast long-distance networks,'' in \emph{Proceedings-IEEE INFOCOM},
  vol.~4.\hskip 1em plus 0.5em minus 0.4em\relax IEEE, 2004, pp. 2514--2524.

\bibitem{quic}
A.~Langley, A.~Riddoch, A.~Wilk, A.~Vicente, C.~Krasic, D.~Zhang, F.~Yang,
  F.~Kouranov, I.~Swett, J.~Iyengar \emph{et~al.}, ``The quic transport
  protocol: Design and internet-scale deployment,'' in \emph{Proceedings of the
  Conference of the ACM Special Interest Group on Data Communication}.\hskip
  1em plus 0.5em minus 0.4em\relax ACM, 2017, pp. 183--196.

\bibitem{remy}
\BIBentryALTinterwordspacing
K.~Winstein and H.~Balakrishnan, ``Tcp ex machina: Computer-generated
  congestion control,'' in \emph{Proceedings of the ACM SIGCOMM 2013 Conference
  on SIGCOMM}.\hskip 1em plus 0.5em minus 0.4em\relax ACM, 2013. [Online].
  Available: \url{http://doi.acm.org/10.1145/2486001.2486020}
\BIBentrySTDinterwordspacing

\bibitem{bbr}
N.~Cardwell, Y.~Cheng, C.~S. Gunn, S.~H. Yeganeh, and V.~Jacobson, ``Bbr:
  Congestion-based congestion control,'' \emph{Queue}, vol.~14, no.~5, p.~50,
  2016.

\bibitem{vegas}
L.~S. Brakmo, S.~W. O'Malley, and L.~L. Peterson, \emph{TCP Vegas: New
  techniques for congestion detection and avoidance}.\hskip 1em plus 0.5em
  minus 0.4em\relax ACM, 1994, vol.~24, no.~4.

\bibitem{p4}
P.~Bosshart, D.~Daly, G.~Gibb, M.~Izzard, N.~McKeown, J.~Rexford,
  C.~Schlesinger, D.~Talayco, A.~Vahdat, G.~Varghese \emph{et~al.}, ``P4:
  Programming protocol-independent packet processors,'' \emph{ACM SIGCOMM
  Computer Communication Review}, vol.~44, no.~3, pp. 87--95, 2014.

\bibitem{sw3}
P.~Bosshart, G.~Gibb, H.-S. Kim, G.~Varghese, N.~McKeown, M.~Izzard, F.~Mujica,
  and M.~Horowitz, ``Forwarding metamorphosis: Fast programmable match-action
  processing in hardware for sdn,'' in \emph{ACM SIGCOMM Computer Communication
  Review}, vol.~43, no.~4.\hskip 1em plus 0.5em minus 0.4em\relax ACM, 2013,
  pp. 99--110.

\bibitem{sw4}
``New cisco asic has a programmable data plane,''
  \url{http://searchnetworking.techtarget.com/news/
  2240177388/New-Cisco-ASIC-has-a-programmable-data-plane}, 2018.

\bibitem{west}
C.~Casetti, M.~Gerla, S.~Mascolo, M.~Y. Sanadidi, and R.~Wang, ``Tcp westwood:
  end-to-end congestion control for wired/wireless networks,'' \emph{Wireless
  Networks}, vol.~8, no.~5, pp. 467--479, 2002.

\bibitem{bb}
T.-Y. Huang \emph{et~al.}, ``A buffer-based approach to rate adaptation:
  Evidence from a large video streaming service,'' \emph{ACM SIGCOMM}, vol.~44,
  no.~4, 2015.

\bibitem{festive}
J.~Jiang \emph{et~al.}, ``Improving fairness, efficiency, and stability in
  http-based adaptive video streaming with festive,'' \emph{IEEE/ACM TON},
  vol.~22, no.~1, 2014.

\bibitem{bola}
K.~Spiteri \emph{et~al.}, ``Bola: Near-optimal bitrate adaptation for online
  videos,'' in \emph{INFOCOM}, 2016, pp. 1--9.

\bibitem{mpc}
X.~Yin \emph{et~al.}, ``A control-theoretic approach for dynamic adaptive video
  streaming over http,'' in \emph{ACM SIGCOMM}, vol.~45, no.~4, 2015.

\bibitem{red}
S.~Floyd and V.~Jacobson, ``Random early detection gateways for congestion
  avoidance,'' \emph{IEEE/ACM Transactions on Networking (ToN)}, vol.~1, no.~4,
  pp. 397--413, 1993.

\bibitem{sred}
T.~J. Ott, T.~Lakshman, and L.~H. Wong, ``Sred: stabilized red,'' in
  \emph{INFOCOM'99. Eighteenth Annual Joint Conference of the IEEE Computer and
  Communications Societies. Proceedings. IEEE}, vol.~3.\hskip 1em plus 0.5em
  minus 0.4em\relax IEEE, 1999, pp. 1346--1355.

\bibitem{rem}
D.~Lapsley and S.~Low, ``Random early marking for internet congestion
  control,'' in \emph{Global Telecommunications Conference, 1999. GLOBECOM'99},
  vol.~3.\hskip 1em plus 0.5em minus 0.4em\relax IEEE, 1999, pp. 1747--1752.

\bibitem{blue}
W.-c. Feng, K.~G. Shin, D.~D. Kandlur, and D.~Saha, ``The blue active queue
  management algorithms,'' \emph{IEEE/ACM transactions on networking}, vol.~10,
  no.~4, 2002.

\bibitem{avq}
S.~Kunniyur and R.~Srikant, ``Analysis and design of an adaptive virtual queue
  (avq) algorithm for active queue management,'' in \emph{ACM SIGCOMM CCR},
  vol.~31, no.~4.\hskip 1em plus 0.5em minus 0.4em\relax ACM, 2001, pp.
  123--134.

\bibitem{pi}
C.~V. Hollot, V.~Misra, D.~Towsley, W.-B. Gong \emph{et~al.}, ``On designing
  improved controllers for aqm routers supporting tcp flows,'' in
  \emph{Proceedings IEEE INFOCOM 2001. Conference on Computer Communications.
  Twentieth Annual Joint Conference of the IEEE Computer and Communications
  Society}.\hskip 1em plus 0.5em minus 0.4em\relax IEEE, 2001, pp. 1726--1734.

\bibitem{gibbens}
R.~J. Gibbens and F.~P. Kelly, ``Distributed connection acceptance control for
  a connectionless network,'' in \emph{COLLOQUIUM DIGEST-IEE}.\hskip 1em plus
  0.5em minus 0.4em\relax Citeseer, 1999, pp. 7--7.

\bibitem{abc}
P.~Goyal, M.~Alizadeh, and H.~Balakrishnan, ``Rethinking congestion control for
  cellular networks,'' in \emph{Proceedings of the 16th ACM Workshop on Hot
  Topics in Networks}.\hskip 1em plus 0.5em minus 0.4em\relax ACM, 2017, pp.
  29--35.

\bibitem{natcp}
S.~Abbasloo, Y.~Xu, H.~J. Chao, H.~Shi, U.~C. Kozat, and Y.~Ye, ``Toward
  optimal performance with network assisted tcp at mobile edge,'' \emph{2nd
  {USENIX} Workshop on Hot Topics in Edge Computing (HotEdge 19)}, 2019.

\bibitem{mec}
(2014) Mobile edge computing introductory technical white paper.
  \url{https://portal.etsi.org/Portals/0/TBpages/MEC/Docs/Mobile-edge_Computing_-_Introductory_Technical_White_Paper_V1\%2018-09-14.pdf}.

\bibitem{reno}
D.~Cox and L.-R. Dependence, ``a review,'' \emph{Statistics: An Appraisal, HA
  David and HT David (Eds.)}, pp. 55--74, 1984.

\bibitem{optimal}
R.~Gail and L.~Kleinrock, ``An invariant property of computer network power,''
  in \emph{Proceedings of the International Conference on Communications},
  Denver, Colorado, June 14-18, 1981, pp. 63.1.1--63.1.5.

\bibitem{jaf}
J.~Jaffe, ``Flow control power is nondecentralizable,'' \emph{IEEE Transactions
  on Communications}, vol.~29, no.~9, pp. 1301--1306, 1981.

\bibitem{ar_vr}
(2016) Immersive vr and ar experiences with mobile broadband.
  \url{http://www.huawei.com/minisite/hwmbbf16/insights/HUAWEI-WHITEPAPER-VR-AR-Final.pdf}.

\bibitem{mahi}
R.~Netravali, A.~Sivaraman, K.~Winstein, S.~Das, A.~Goyal, and H.~Balakrishnan,
  ``Mahimahi: A lightweight toolkit for reproducible web measurement,'' 2014.

\bibitem{pens}
H.~Mao \emph{et~al.}, ``Neural adaptive video streaming with pensieve,'' in
  \emph{ACM SIGCOMM}, 2017.

\bibitem{3gpp}
(2018) Policy and charging control architecture (3gpp ts 23.203 version 14.6.0
  release 14).
  \url{https://www.etsi.org/deliver/etsi_ts/123200_123299/123203/14.06.00_60/ts_123203v140600p.pdf}.

\bibitem{DiffServ}
S.~Blake, D.~Black, M.~Carlson, E.~Davies, Z.~Wang, and W.~Weiss, ``An
  architecture for differentiated services,'' Tech. Rep., 1998.

\end{thebibliography}

\end{document}